\documentclass{jfm}
\usepackage{color}
\usepackage{amsmath}
\begin{document}

\newtheorem{lemma}{Lemma}
\newtheorem{corollary}{Corollary}
\newcommand{\be}{\begin{equation}}
\newcommand{\ee}{\end{equation}}
\newcommand{\bea}{\begin{eqnarray}}
\newcommand{\eea}{\end{eqnarray}}
\newcommand\cpc[1]{{\color{black}#1}}
\newcommand\cpcf[1]{{\color{black}#1}}
\newcommand\mha[1]{{\color{black}#1}}
\shorttitle{Goldilocks ocean mixing} 
\shortauthor{A. Mashayek, C. P. Caulfield \& M. H. Alford} 

\title{Goldilocks mixing in \color{black}oceanic \color{black} shear-induced turbulent overturns}
\author
 {
 A. Mashayek\aff{1}, C.P. Caulfield\aff{2,3}, M. H. Alford\aff{4}
  \corresp{\email{mashayek@ic.ac.uk}},
  }

\affiliation
{
\aff{1}
Imperial College London, London, UK
\aff{2}
BP Institute for Multiphase Flow, University of Cambridge, Madingley Rise, Madingley Road, Cambridge CB3 0EZ, UK
\aff{3}Department of Applied Mathematics and Theoretical Physics, University of Cambridge,
Wilberforce Rd, Cambridge CB3 0WA, UK
\aff{4}
Scripps Institution of Oceanography, University of California San Diego, La Jolla, USA
}
\maketitle

\begin{abstract}
We present a new physically-motivated parameterization, based on the ratio of Thorpe and Ozmidov scales,  for the irreversible turbulent flux coefficient $\Gamma_{\mathcal M}= {\mathcal M}/\epsilon$, i.e. the ratio of the irreversible rate ${\mathcal M}$ at which the background potential energy increases in a stratified flow due to macroscopic motions to the dissipation rate of turbulent kinetic energy. Our parameterization covers all three key phases \cpc{(crucially, in time)} of a shear-induced stratified turbulence life cycle: the initial, `hot' growing phase, the intermediate energetically forced phase, and the final \cpc{`cold'} fossilization decaying phase. Covering all three phases allows us to highlight the importance of the intermediate one, to which we refer  as the `Goldilocks' phase due to its apparently optimal (and so neither too hot nor too cold, but just right) balance, in which energy transfer from background shear to the turbulent mixing is most efficient. $\Gamma_{\mathcal M}$ is close to 1/3 during this phase, which we demonstrate \cpc{appears to be} related to an adjustment towards a critical \cpcf{or marginal} Richardson number for sustained turbulence  $\sim 0.2-0.25$.
\cpcf{Importantly, although buoyancy  effects are still significant at leading order for the turbulent dynamics during this intermediate phase,  the marginal balance in the flow ensures that the turbulent mixing of the (density) scalar is nevertheless effectively `slaved' to the turbulent mixing of momentum.} We present supporting evidence for our parameterization through comparison with six oceanographic datasets that span various turbulence generation regimes and a wide range of geographical location and depth. Using these observations, we highlight the crucial significance of parameterizing an inherently variable flux coefficient for capturing the turbulent flux associated with  rare energetic, yet fundamentally shear-driven \cpcf{(and so crucially not strongly stratified)} overturns that make a disproportionate contribution to the total mixing. 
\color{black} We also highlight the importance of representation of young turbulent patches in the parameterization for connecting the small scale physics to larger scale applications of mixing such as ocean circulation and tracer budgets\color{black}.
Shear-induced 
turbulence is therefore central to irreversible mixing in the world's oceans, apparently even close to the seafloor, \cpc{and it is critically important to appreciate the inherent time dependence and evolution of mixing events: history matters to mixing.} 
\end{abstract}

\section{Introduction}
\mha{The} efficiency of shear-induced turbulent mixing has been widely studied over the past several decades using various observational, experimental, theoretical and computational approaches \citep{PC03,ivey,Gregg2018MixingOcean,Caulfield2020,Caulfield2021LayeringFlows}. Many works have relied on parameterizing the turbulent flux coefficient, $\Gamma$, (loosely the ratio of the mixing rate, with various definitions, to the turbulent kinetic energy dissipation rate $\epsilon$) in terms of various non-dimensional parameters such as the buoyancy Reynolds number, $Re_b=\epsilon/(\nu N^2)$, where $\nu$ is the kinematic
viscosity and $N^2$ is an appropriate buoyancy frequency, 
\cpc{the turbulent Froude number $Fr= \epsilon/[N k]$ where $k$ is the turbulent kinetic energy (density)}
or (for sheared turbulence) the Richardson number, $Ri=N^2/S^2$, where $S$ is the vertical shear of some `background' streamwise velocity. Indeed, some of these approaches are dimensionally insufficient to represent mixing as previously discussed in the literature \citep[e.g. see][]{Ivey1991650,Shih2005,Mashayek2011TurbulenceMechanism,MP13,mater2014unifying}. More specifically, it has been shown that despite an empirically-observed emergence of power-law dependence of $\Gamma$ on $Re_b$ over various turbulent regimes, when observational, experimental, and numerical data of $\Gamma$ are plotted against $Re_b$, they typically do not overlap, but rather display a wide  scatter \citep{bouffard2013diapycnal,mashayek2017efficiency,monismith2018mixing}. It has been demonstrated that such scatter has leading order implications for the role of mixing in sustaining the deep branch of ocean circulation \citep{de2016impact,mashayek2017efficiency,Cimoli2019SensitivityEfficiency}. In fact, recent observational work of \citet{Ijichi2020HowAbyss} casts further doubt on parameterizing $\Gamma$  in terms of $Re_b$ alone. 

Of course, such nondimensional parameters can be interpreted as ratios of length scales, and  an alternative but thus clearly related approach to the parameterization  of $\Gamma$, \color{black} first proposed by \citet{Ivey1991650}\color{black}, has been to consider the ratio of the Ozmidov and Thorpe scales, $R_{OT}$, defined as
\begin{equation}
    R_{OT}\equiv L_O / L_T, \label{eq:rotdef}
\end{equation} where 
$L_O\equiv (\epsilon/N^3)^{1/2}$ and the Thorpe scale is the rms of the displacements required to reorder a particular profile of density measurements into a monotonically decreasing profile. Use of $R_{OT}$ has been suggested by some as an indication of `overturn age' and therefore a suitable basis for a parameterization of $\Gamma$ \citep{Ivey1991650,smyth_etal_2001,mashayek2017role,Ijichi2018ObservedOcean}. However, for practical reasons, in  physical oceanography it is often operationally assumed that $R_{OT}$ is equal or close to one to infer estimates of rate of dissipation of kinetic energy. 

At its heart,
this assumption relies on the combination of the ideas that $L_T$ may be thought of as the characteristic scale of an overturning turbulent event, while $L_O$ is the largest (vertical) scale which is not strongly \cpcf{or (perhaps more accurately) dominantly} affected by the background stratification, and that these scales should be similar for a vigorous turbulent patch.
\cpc{However, this view does not capture the importance of the inherent time-dependence of mixing events, and in particular that mixing events evolve through a life cycle, with subsequent phases retaining an imprint of previous stages in the flow evolution. As \cite{Villermaux2019} noted for passive scalar mixing, followed up by  \cite{Caulfield2021LayeringFlows} in the density-stratified context, history really does matter for mixing, and so it is important to parameterize mixing events throughout their entire life cycle. }

In this work we build on more fundamental theoretical grounds, specifically the assumption of \cpc{the} existence of an inertial subrange, mixing length theory, and criteria for shear instability, and propose a  parameterization for $\Gamma$ in terms of $R_{OT}$. \cpc{Our approach thus has two central pillars. First, as we suppose shear instability drives the mixing, the flows always have  a stratification that is \cpcf{not too strong to suppress the growth of an instability 
to a significant (overturning) amplitude. Here, we refer to such a stratification as {\it subcritical}, in the specific sense
that the key parameter quantifying the relative strength of the stratification to the shear, the Richardson number (defined more precisely below) is sufficiently small to allow
the growth of such a shear-driven overturning instability. It is always important to remember that buoyancy  can still play a leading order role in such subcritical flows, and as we discuss further below, there is accumulating evidence that flows generically adjust so that characteristic values of the Richardson number become close to the critical or marginal value at which instability can (just) occur.} Second, the inherent time-dependence of such mixing is a distinguishing characteristic which must always be captured by our modelling and/or parameterization.}  

\color{black} In the limit $R_{OT} \gg 1$, the parameterization reduces to a scaling relation previously offered by others, \cpc{however our inherently time-dependent yet \cpcf{subcritically stratified} interpretation relies on understanding this limit as being associated with the late-time decay of an `old' shear-driven overturning mixing event.} In the $R_{OT} \ll 1$ limit, however, the parameterization reduces to a new scaling for the (crucially) `young' turbulent patches. We argue that neither of the two limits' scaling is correct \cpc{in isolation}, but that their merged form is appropriate since ocean data \cpc{are} known to be \mha{distributed near} $R_{OT} \sim 1$ \citep[][ also  shown in \S \ref{sec:data} ]{Dillon1982VerticalScales,ferron1998mixing,Thorpe_book}. 
\cpc{Vitally, this intermediate value should still be associated with  a \cpcf{subcritical, shear-driven overturning} mixing event at an intermediate, and energetic stage in its temporal evolution.}
We also determine the value of the sole coefficient in the parameterization based on the \cpc{further, physically and empirically motivated} assumptions of turbulent Prandtl number $\sim 1$ and a critical Richardson number \cpc{around} $1/4$. 
\cpc{These two assumptions are of course at their heart assumptions that the stratification is} \cpcf{not too strong to dominate all aspects of the dynamics}: 
\cpcf{the turbulent Prandtl number being around one implies that  scalar
mixing is (at least approximately) slaved to the mixing of momentum; while 
the very presence of shear-driven instabilities requires a sufficiently low value of the Richardson number.}

To support this proposed parameterization,  we demonstrate  good agreement with several oceanic datasets corresponding to various turbulence types and different depth ranges. The value inferred for the coefficient of the parameterization based on data regression closely matches the theoretical prediction.
Thus, our primary finding  is that a parameterization based entirely on physical grounds (even up to its sole coefficient) appears to explain the efficiency of mixing of the observed turbulent patches for a range of oceanic turbulent processes, as long as, crucially, the source of the turbulence is background shear, \cpc{and the conceptual picture is that the turbulence is time-developing, evolving through various phases of shear-driven mixing events, inevitably associated with stratification \cpcf{not being too strong}.} 

Our theory and observations suggest $\Gamma \sim 1/4-1/3$ for most of such shear-driven turbulent data when $R_{OT}$ is within a factor of 3 of unity. \cpc{Significantly,  the mixing coefficient} $\Gamma$, \cpc{is predicted to be} slightly larger, yet still quite comparable to, the classical  value of $0.2$. \cpc{This somewhat more efficient mixing is associated with} a phase of energetic turbulence in which the stratification, not yet overly eroded, gives rise to a rich cascade of hydrodynamic instabilities that efficiently channel energy from the parent overturn and the background shear through shear production. \cpc{Through analogy with the common usage in astrobiology of `Goldilocks zone' for the circumstellar habitable zone,} we refer to this phase as  {\it Goldilocks} turbulence, \cpc{as it is neither too hot, nor too cold, but just right.  

Less informally, it demonstrates the vital importance of appreciating the time history of shear-driven mixing events, specifically the ensuing efficient mixing occurring after the break down of a relatively large primary shear-driven overturning, which once again, can only arise if the stratification may be considered to be \cpcf{below some critical strength}.} We also show evidence, based on numerical simulations, that for energetic ocean turbulence (i.e. at sufficiently high Reynolds numbers and \cpc{sufficiently small} Richardson numbers) most of the overturn evolution life cycle corresponds to the Goldilocks phase ($R_{OT}\sim 1$), thereby giving credence to arguments of \citet{gregg1987diapycnal} and \citet{Caldwell1983OceanicCreation} as opposed to \citep{gibson1987fossil,Baker1987SamplingTurbulence} who argued that most observations were \cpc{what they referred to as {\it fossilized} turbulence}, \cpc{i.e. the turbulence from late in the life cycle of a mixing event}.

Finally, although it is undoubtedly tempting to use an averaged value of  $\Gamma$ in models, we demonstrate that doing so  \cpc{has the potential to} lead to large inaccuracies, due \cpc{perhaps unsurprisingly} to a range of inherent nonlinearities in the system. For example, we show that the total buoyancy flux obtained through summation of \cpcf{the} contribution of  individual patches in various datasets \cpc{is} strongly dominated by those patches with the largest values of \cpcf{the} rate of dissipation of kinetic energy (as one would expect) and so it is important to capture \cpc{as accurate a value as possible} of $\Gamma$ for such patches, rather than using a mean value extracted from data associated with all patches. We also highlight the significance of the large $\Gamma$ associated with young turbulence for quantification of bulk mixing in a grid cell of a coarse resolution climate model.

To present our key points, the rest of the paper is organized as follows. In \S \ref{Basic_Definitions}, we present definitions of the various important length scales and parameters. We then describe the key aspects of the phenomenology of a shear-driven mixing event in \S \ref{phenom}, extracted from a numerical simulation. 
In \S \ref{sec:data}, we briefly describe six oceanic datasets, which we then use to verify our proposed parameterization for $\Gamma$ in terms of $R_{OT}$ in \S \ref{ROT}. \cpc{We also compare and contrast this parameterization with previous studies, in particular those of \cite{Maffioli2016MixingTurbulence} and \cite{Garanaik2019OnFlows}  based around the use of the turbulent Froude number.  Specifically, we highlight the interesting fact that similar scalings can arise based on conceptually different physical interpretations, not relying on the inherent time-dependence and weak stratification at the heart of our interpretation of shear-driven overturning mixing events.} We highlight three key implications of our parameterization in \S \ref{3points}. In \S \ref{sec:Reb} we argue that while observationally desirable, parameterization of $\Gamma$ based on $Re_b$ or $(Re_b,Ri)$ might be impractical. We draw our conclusions in \S \ref{sec:conc}, and suggest some future avenues of research. 
 
\color{black}
\section{Basic Definitions}
\label{Basic_Definitions}
The basic turbulence scales that we employ herein are the Kolmogorov scale, $L_K$, representing the scale below which viscous dissipation takes kinetic energy out of the system, the Ozmidov scale, $L_O$, the maximum (vertical) scale that is not strongly affected by stratification, the Corrsin scale, $L_C$, the maximum scale that is not strongly affected by the background shear, and finally the Thorpe scale, $L_T$, a geometrical vertical scale characteristic of  displacement of notional fluid parcels within an overturning turbulent patch. The first three may be defined as
\begin{equation}
    L_K=\left(\frac{\nu^3}{\epsilon}\right)^\frac{1}{4}, \qquad L_O=\left(\frac{\epsilon}{N^3}\right)^\frac{1}{2}, \qquad
    L_C=\left(\frac{\epsilon}{S^3}\right)^\frac{1}{2},\label{eq:lengthsdef}
\end{equation}
where $\nu$ is the kinematic viscosity of the fluid, $\epsilon$ is here the  rate of dissipation of  turbulent kinetic energy, $N$ is the characteristic buoyancy frequency and $S$ is the background shear. It is  important to remember that the calculation of characteristic values of $\epsilon$, $N$ and $S$ for a given flow 
must always involve some averaging, whether over ensembles, space or time.  

Note also that while we conveniently will refer to $L_T$ as a turbulence length scale, it is more appropriate to think of it as a geometric, and somewhat subjective property of  an identified  `patch' of turbulence. \citep{Thorpe_book,Dillon1982VerticalScales,Mater2015BiasesData,chalamalla2015mixing,mashayek2017role}. Nevertheless, it has proved to be an important measure which can be readily inferred from observations.
Once such length scales have been defined, their ratios lead naturally to various non-dimensional parameters. In particular, definitions of a \mha{Richardson number and a} buoyancy Reynolds number can be written as:
\begin{equation}
Ri = \left ( \frac{L_C}{L_O} \right )^{2/3}=\frac{N^2}{S^2}, \qquad
Re_b=\left( \frac{L_O}{L_K} \right )^{4/3}=\frac{\epsilon}{\nu N^2},
\label{Ri_Reb}
\end{equation}
where here we further assume that the background shear $S$ is the vertical shear of horizontal velocity.

In this paper, we are interested in parameterizing various properties of turbulent mixing, in particular its `efficiency'. Here, we choose to  define the {\bf instantaneous}  efficiency of turbulent mixing in terms of a ratio of conversion rates: i.e. the `mixing' rate ${\mathcal M}$ at which the minimal background potential 
energy is irreversibly 
increasing due to macroscopic fluid motions \citep{Winters,CP00} divided by ${\mathcal M}+ \epsilon$, the rate at which kinetic energy is being irreversibly \mha{lost}:
\begin{equation}
\mathcal{E}_i=\frac{\mathcal{M}}{(\mathcal{M}+{\epsilon})}, 
\end{equation}
As expected for an `efficiency', ${\mathcal E}_i \leq 1$ strictly. A corresponding flux coefficient (often somewhat confusingly referred to as an efficiency, although in principle ${\mathcal M} > \epsilon$ is possible) may then be defined as 
 \begin{equation}
 \Gamma_{\mathcal M}=\frac{\mathcal{E}_i}{1-\mathcal{E}_i}=\frac{\mathcal{M}}{\epsilon},
 \end{equation}
 where the subscript ${\mathcal M}$ makes explicit that this definition of the turbulent flux coefficient is in terms of the irreversible mixing rate. \cpc{(As discussed in recent reviews by \cite{Gregg2018MixingOcean,Caulfield2020,Caulfield2021LayeringFlows}, there are several different possible definitions for this flux coefficient, and it is important to be clear which particular definition is being utilised when comparing different data sources.)}
 
 Generically, there is no reason to suppose that $\Gamma_{\mathcal M}$ is constant. As originally argued by \cite{Osborn_1980}, an
 understanding of the properties of $\Gamma_{\mathcal M}$ can lead to  a parameterization for the (vertical) eddy or turbulent diffusivity of buoyancy ${\kappa}_T$, defined as 
 \begin{equation}
\kappa_T \equiv -\frac{\mathcal{B}}{N^2}=-\frac{<w'b'>}{N^2}
\label{eq:Kdef}
 \end{equation}
 where $w'$ and $b'$ are turbulent velocity and buoyancy perturbations, and hence ${\mathcal B}$ is an (appropriately
 averaged) vertical buoyancy flux. Assuming that transport terms, reversible processes and spatio-temporal variability in general can be ignored (see, for example \cite{MCP13} for further discussion) ${\mathcal B}$ can be approximated by ${\mathcal M}$ \cpc{in many situations of interest}, and so the enhanced turbulent diffusivity is given by
 \begin{equation}
   \frac{\kappa_T}{\kappa} \equiv  -\frac{\mathcal{B}}{\kappa N^2} 
   \approx \Gamma_{\mathcal M} \left( \frac{\nu}{\kappa} \right ) \left ( \frac{\epsilon}{\nu N^2} \right ) = \Gamma_{\mathcal M} Pr Re_b ,\label{eq:prdef}
 \end{equation}
where $\kappa$ is the (molecular) diffusivity of the buoyancy field, and so $Pr$ is a (molecular) Prandtl number.

\section{Phenomenology of shear-driven mixing events}
\label{phenom}
\begin{figure}
\begin{center}
\includegraphics[trim=0 0 0 0,clip,width=1\textwidth]{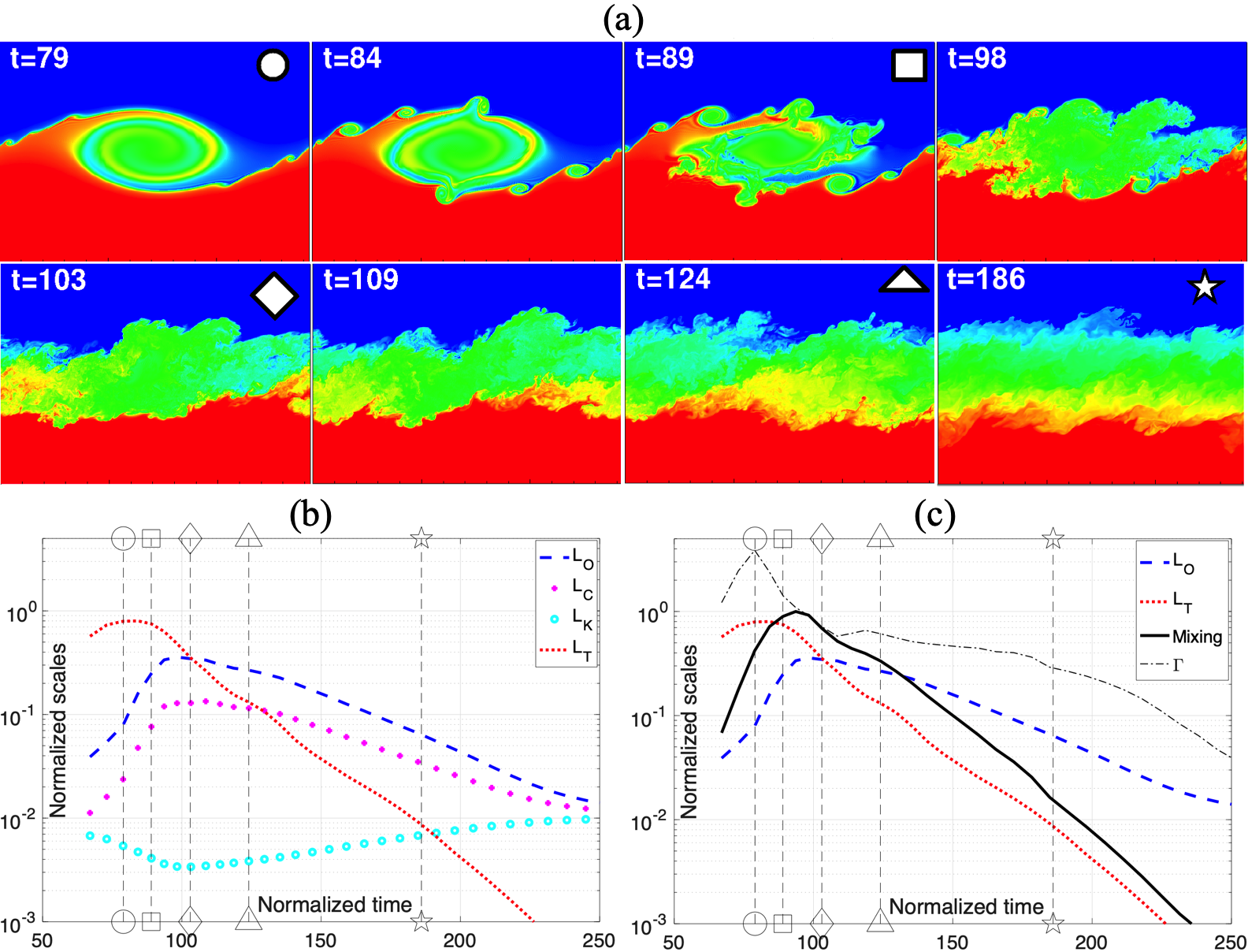}
\caption{(a) Turbulent life  cycle of a shear-driven Kelvin-Helmholtz instability. Colours represent density, with red, green and blue representing dense, intermediate and light waters, respectively. (b) Evolution of the various (nondimensional) length scales defined in section \ref{Basic_Definitions}, scaled with the time-evolving half-depth of the shear layer \citep[following][]{mashayek2017role}. (c) Evolution of $L_T$ and $/L_O$ as well as the 
nondimensional irreversible mixing rate $\tilde{\mathcal{M}}$
(scaled with $U_0^3/d_0$) and turbulent flux coefficient $\Gamma_{\mathcal M}$.  The vertical dashed lines in panels (b,c) mark various characteristic times that correspond to different sub-panels in panel (a) with similar symbol markings. These plots are reproduced from the data associated with a simulation with $Re_0=6000$, $Ri_b=0.16$, $Pr=1$, $R=1$ from \cite{MCP13}.}
\label{Fig1}
\end{center}
\end{figure}

For the construction of a physically motivated mixing parameterization, it is useful to consider the time evolution of various length scales and the flux coefficient during the turbulence life cycle of a transient, shear-driven mixing event, as a particularly simple model of a breaking wave. To this end, in figure \ref{Fig1} we describe the evolution of turbulence due to the breakdown of the canonical shear instability referred to as the Kelvin-Helmholtz instability (KHI). The KHI paradigm has been commonly suggested to be relevant to oceanic turbulent overturning mixing events \citep{smyth_etal_2001,Mater2015BiasesData}\mha{;} we will provide further support towards this idea throughout this paper. 

Panel (a) illustrates growth of a Kelvin-Helmholtz `billow', its turbulence transition, break down, and relaminarization, from a simulation of a flow with initial hyperbolic tangent velocity and buoyancy distributions:
\begin{eqnarray}
{\mathbf u} &=& U(\tilde{z}) \hat{\mathbf x};
\ U(\tilde{z}) = \frac{U_0} \tanh \left ( \frac{z}{d_0} \right ) ; \ 
b= b_0  \tanh \left ( \frac{z}{\delta_0} \right ),
\label{eq:udef}\\
Re_0 &\equiv& \frac{U_0 d_0}{\nu};
\ Ri_b \equiv \frac{b_0 d_0}{U_0^2} ; \ R \equiv \frac{d_0}{\delta_0}, 
\label{eq:bdef}
\end{eqnarray}
defining the appropriate flow Reynolds number $Re_0$,
`bulk' Richardson number $Ri_b$ and scale ratio $R$, $\hat{\mathbf x}$ is the unit vector in the streamwise $x-$direction, and tildes denote dimensionless quantities.  For this particular simulation (see \cite{MCP13} for more details) $Re_0=6000$, $Ri_b=0.16$, $R=1$ and $Pr=1$.
\cpcf{In the sense discussed in the Introduction, this stratification
is definitely {\it subcritical}, in that shear-driven overturning instabilities are able to grow to finite amplitude, and indeed to break down to turbulence.}

Physical and quantitative interpretation of the associated mixing may be achieved through consideration of the various turbulence scales shown in panel (b). 
There are three key phases \cpc{which occur sequentially during the life cycle of a generic shear-driven mixing event}: an early growing young or `hot' phase; an intermediate, energetic and efficient mixing phase; and an ultimate decaying or `cold' phase. 
As we demonstrate, the intermediate phase appears to lead to `optimal' mixing, neither too hot nor too cold but `just right', so  \cpc{as already mentioned in the Introduction}, we refer to this as the `Goldilocks mixing' phase, inspired by the fairy tale.
In what follows we will describe these three phases by discussing the evolution of the relevant above-defined turbulent length scales. The specific methodology for calculating of the scales from DNS were discussed in \cite{mashayek2017role}.

By dimensionless time $\tilde{t}=79$ (scaled with the advective timescale $d_0/U_0$), the primary overturn has fully grown, as is marked by the peaking of $L_T$. 
This represents an accumulation of available potential energy (APE) from the Kinetic Energy (KE) reservoir, originally stored in  the background shear. Upon the saturation of the primary billow, small scale turbulence grows, effectively through feeding on the APE. This leads to a significant increase in the rate of dissipation of turbulent kinetic energy, leading to  a relatively rapid growth of the Ozmidov and Corrsin scales and a decrease in the Kolmogorov scale, together representing a widening of the inertial subrange within which energy can be transferred from the injection scale to dissipation scale through a cascade of eddies, largely unaffected by either stratification or shear. 

A critical time in the turbulence life cycle is when the characteristic scales of the turbulence approach those associated
with the initial vertical scale of the primary overturn, i.e.
when 
$L_O\sim L_T$. Under the (reasonable) assumption that
it is appropriate to think of $L_T$ as the injection scale
for the turbulent motions, due to the implied conversion of APE to KE at this scale, then this critical time marks the
instant at which the broadest possible range of scales largely unaffected by stratification occurs, and so there is an opportunity for efficient extraction of energy from the background \color{black}\citep[see][and references therein for a discussion]{mashayek2017role}\color{black}. Unsurprisingly, $L_C < L_O$, as it is to be expected that the characteristic Richardson number, defined in (\ref{Ri_Reb}) in terms of the length scales, $Ri < 1$. Therefore, although the scales below $L_O$ are largely unaffected by the stratification, there is typically
a range of scales still strongly affected by the background shear, which nevertheless appear to still allow efficient
mixing processes to occur \cpc{where it is appropriate to think of the (dynamical) influence of the stratification as being \cpcf{perhaps important, but definitely not dominant.} }

After this critical time,  APE is depleted rapidly, as manifested through the relatively sharp drop in $L_T$.  $L_O$ and $L_C$ also decay,  albeit markedly more slowly than $L_T$. We can interpret this behaviour  through remembering that $L_O$ and $L_C$  represent the largest  scales that the turbulent eddies notionally could have accessed had they been sufficiently `fed' energetically. However, since
the turbulence has decayed so much,  the overturning (and presumably injection) scale $L_T$ has become smaller  than both $L_O$ and  $L_C$. This empirically observed decay lag between $L_O$ and $L_T$ is of significance for the theoretical arguments that we  make in the following sections. \cpc{Crucially, this decay should not be interpreted as anisotropic turbulence collapse due to `strong' stratification, but rather primarily \cpcf{due} to kinetic energetic losses to viscous dissipation, and, to a lesser extent, to irreversible mixing increasing the potential energy of the system.} Ultimately, the flow enters a (molecular) diffusion-dominated regime, as diffusivity drops back to molecular values and so $L_O\rightarrow L_K$, as is shown as $t\rightarrow 250$ in panel (b).

Panel (c) shows the evolution of the turbulent flux coefficient, $\Gamma_{\mathcal M}$, and the (nondimensional) irreversible mixing rate  $\tilde{\mathcal{M}}$ \cpc{i.e. scaled with the characteristic scales $U_0^3/d_0$}. As would be expected based on the evolution of scales in panel (b),  mixing is most efficient between the peak of $L_T$ (and so when APE is largest) and the `critical' time when $L_T\approx L_O$. This reflects the richness of the zoo of hydrodynamic instabilities within the broad inertial subrange and their ability to stir and hence irreversibly mix density (and tracer) gradients  efficiently (see for example \cite{MP1,MP2} for a detailed analysis of the link between the cascade of hydrodynamic instabilities that facilitate turbulence breakdown and the characteristics of turbulent mixing). The subsequent decay of the mixing rate is highly correlated with the  decay of its \cpc{(driving)} source of energy (i.e. APE), and so is highly correlated with the decay of $L_T$.  
From the time APE saturates onwards, $L_T/L_O$ decays \cpc{essentially}  monotonically since $L_T$ consistently decays (relatively rapidly) and $L_O$ first grows and then decays at a slower rate. For this reason, and as we will discuss further, this ratio  has been suggested to be a good proxy for evolution of mixing during turbulent life cycles in shear flows \cite{smyth_etal_2001,mashayek2017role}. This is a reasonable
suggestion, as is apparent from the \cpc{apparent} correlation  between \cpc{the ratio} $L_T/L_O$ and $\Gamma_{\mathcal M}$.

\section{Data}\label{sec:data}
We now consider six oceanic datasets to constrain and inform our proposed parameterizations for $\Gamma$, \cpc{principally} in terms of $R_{OT}$, and then \cpc{briefly in terms of} $Re_b$.  All six were collected from free-falling microstructure profilers, which sample shear and temperature with sub-centimeter resolution in order to estimate the dissipation rate of kinetic energy, $\epsilon$, and the thermal variance dissipation rate. 
Each also carries a conductivity-temperature-depth (CTD) instrument in order to sample the profiles of temperature and salinity needed to estimate potential density and buoyancy frequency $N$.  Ozmidov scales are then directly calculated from $\epsilon$ and $N$,  \cpc{while} Thorpe scales are calculated by re-ordering the measured potential density, and evaluating the rms of the required displacements of individual measurements, which is essentially the one-dimensional observational analogue of the approach used by \cite{Winters}.  

Following the methodology proposed by \cite{moum1996efficiency}, a turbulent flux coefficient $\Gamma_\chi$ is computed from $\epsilon$ and an appropriately scaled version of \cpc{the thermal dissipation rate corresponding to the destruction rate of  buoyancy variance, or equivalently, under the further assumption that the characteristic buoyancy frequency is constant, the destruction rate of available potential energy}:
\begin{equation}
    \Gamma_\chi \equiv \frac{\chi}{\epsilon}. \label{eq:gammachi}
\end{equation}
As discussed further in recent reviews \citep{Gregg2018MixingOcean,Caulfield2020,Caulfield2021LayeringFlows}, under certain circumstances it is 
reasonable to suppose that $\Gamma_{\mathcal M}\simeq \Gamma_\chi$, an assumption we make here when comparing observational and numerical simulation data.

The Tropical Instability Wave Experiment (TIWE) dataset includes  turbulent patches sampled at the equator at 140$^o$W in the shear-dominated upper-equatorial thermocline, between 60m and 200m depths, spanning both the upper and lower flanks of the Pacific Equatorial Undercurrent \citep{lien1995turbulence,smyth_etal_2001}. The FLUX STAT (FLX91) experiment sampled turbulence at the thermocline ($\sim$350-500m depth), in part generated through shear arising from downward-propagating near-inertial waves, about 1000 km off the coast of northern California \citep{moum1996efficiency,smyth_etal_2001}.  The IH18 experiment measured full-depth turbulence (up to $\sim$5300m deep) primarily generated by tidal flow over the Izu-Ogasawara Ridge (western Pacific, south of Japan), a prominent generation site of the semidiurnal internal tide that spans the critical latitude of 28.88N for parametric subharmonic instability \citep{Ijichi2018ObservedOcean}. The Samoan Passage data are measurements of abyssal turbulence generated by hydraulically-controlled flow over sills in the depth range 4500-5500m in the Samoan Passage, an important topographic constriction in the deep limb of the Pacific Meridional Overturning Circulation \citep[see][we use data from the latter]{alford2013turbulent,Carter2019APassage}. The BBTRE data are from turbulence induced by internal tide shear in the deep Brazil Basin ($\sim$2500-5000m depth) and were acquired as a part of the original Brazil Basin Tracer Release Expermient (BBTRE; \cite{polzin1997spatial}), recently re-analyzed by \cite{Ijichi2020HowAbyss}. Also re-analyzed by \cite{Ijichi2020HowAbyss}, we use the data from DoMORE which focused on flow over a sill on a canyon floor in the Brazil Basin \citep{Clement2017TurbulentRidge,Ijichi2020HowAbyss}. In total, these datasets comprise over 36,000 patches and six different turbulence generation regimes over a wide range of depth and geographical locations.

\section{Parameterizing $\Gamma_{\mathcal M}$ as a function of $L_O/L_T$}
\label{ROT}
Building on the phenomenology of the overturning  life cycle  illustrated in figure \ref{Fig1},  in this section we propose a  functional dependence of $\Gamma_{\mathcal M}$ on the key ratio of length scales, which has historically been defined as $R_{OT}  \equiv L_O/L_T$, as in equation (\ref{eq:rotdef}).
Although the canonical example shown in figure 1 was of the break-down of a KHI, 
 for our following arguments to hold,  all that is needed is the notion of an overturning-induced turbulence (not necessarily a \cpc{classical} shear instability) that comprises a growing phase, an energetically mixing phase, and a decaying   phase.  As we argue in more detail below, this decaying phase can be considered, at least in some sense, as a `fossilizing' phase of turbulence, with some points of analogy with the classical arguments of \cite{gibson1987fossil}. 
 Importantly, underlying  this approach is the concept the 
 stratification is \cpcf{is not too strong, so that in can be considered to be subcritical, in the very specific} sense that the developing turbulence is not itself \cpcf{thoroughly} dominated by the stratification, but rather that vigorous
 overturnings have managed to develop. This concept
 can be supported by identification of spatio-temporally localised
 turbulent patches within flows \cpc{where the stratification, \cpcf{though possibly dynamically important, is not, at least locally, dominant}, while}  in bulk terms, \cpc{the larger scale flow} can be classified as being strongly stratified, as demonstrated by \cite{Portwood2016}.
 
 \cpc{\subsection{Length scale ratio for $\Gamma_{\cal M}$}\label{sec:fund}}
 First, we assume that the largest energy containing eddies of length $L_i$ inject energy into the system and that such an input is ultimately balanced by an energy sink at the viscous dissipation scale after a (net) forward cascade \citep{Kolmogorov1941TheNumbers,Taylor1935StatisticalStream}. This leads to the classic turbulent (and inherently unstratified) integral scale, 
\begin{equation}\label{intScale}
L_i \propto \frac{k^{3/2}}{\epsilon},
\end{equation}
where $k$ is the turbulent kinetic energy (TKE). 
This scaling has at its heart that the turbulence being considered is not strongly affected by stratification.

Next we invoke classic mixing length theory \citep{Prandtl1925REPORTTURBULENCE,Taylor1915EddyAtmosphere}, which gives the turbulent diffusivity in terms of a mixing length $L_{m}$ and \cpc{$k$}:
\begin{equation}\label{mixing_length}
\kappa_T\propto L_{m}\ k^{1/2}.
\end{equation}
Here  \cpcf{the (possibly significant) dynamical} influence of ambient stratification 
may be embedded within the mixing length, but it  also contains an implicit assumption 
that the mixing of 
the \cpcf{buoyancy} is assumed to be set by \cpcf{(and so in a sense slaved to)} the mixing
of momentum, as quantified by the right-hand side of the expression. Equivalently, it is assumed that the 
turbulent Prandtl number 
$Pr_T \sim O(1)$, defined as 
\begin{equation}
    Pr_T \equiv \frac{\nu_T}{\kappa_T};\ 
    \nu_T \equiv \frac{\left \langle u' w' \right \rangle}{ S}\simeq \frac{P}{S^2},
    \label{eq:nutdef}
\end{equation}
where ${\nu}_T$ is the eddy diffusivity (of momentum), and we assume that the turbulent production $P$ is dominated by Reynolds stress extraction from the mean shear $S$.
\cpcf{Therefore, it is most definitely not appropriate to think of the flow as `strongly' stratified in any meaningful sense.}

Also, by its very nature `overturning' mixing must be occurring in (at least locally) \cpcf{a subcritical} stratification as observed by \cite{AlfordPinkel00}.
\cpcf{Such overturning mixing is qualitatively different from} the `scouring' mixing in the vicinity of `sharp' and strong density interfaces, a classification distinction drawn by \cite{Woods2010}
(see  \cite{Caulfield2020,Caulfield2021LayeringFlows} for more discussion).

Finally, we invoke the Osborn balance presented previously in  (\ref{eq:prdef}),
\begin{equation}\label{Osb}
\kappa_T\approx \Gamma_{\mathcal M} \frac{\epsilon}{N^2}.
\end{equation}
These three ingredients can form the basis for a simple parameterization that fits all the data introduced earlier, \cpc{provided various data are thought of as being sampled at different stages in the time evolution of a particular mixing event.} To achieve this, however, it is \cpc{also} important to distinguish between the two characteristic length scales defined in (\ref{intScale}) and (\ref{mixing_length}).  
Remembering that $\epsilon$ and $N$ can in turn be related to the Ozmidov length scale since $L_O^2=\epsilon/N^3$,
it is possible to use these equations to arrive at
\begin{equation}\label{2Lbalance}
\Gamma_{\mathcal M}\propto \left[\frac{L_i L_{m}^3}{L_O^4} \right]^{\frac{1}{3}}.
\end{equation}

To develop a parameterization, it is now necessary to identify appropriate candidate lengths  for $L_m$ and $L_i$ that can actually be quantified.
Since here we are interested
in overturns far from boundaries, we expect
that the distance to the boundary is not a relevant candidate. Motivated by the phenomenology discussed
in \S \ref{phenom}, we argue that there are different appropriate candidates for these scales at different stages of the flow evolution of an overturn \cpc{in a sufficiently weakly stratified flow.}

\cpc{\subsection{`Young' turbulence scaling for $\Gamma_{\cal M}$}\label{sec:young}}
During the first growing phase of the turbulence,  $L_T \gg L_O$. During that phase, it is appropriate to identify the mixing length $L_m \sim L_T$, as the Thorpe scale is  the relevant scale over which the fluid gets stirred and mixed against the background stratification. 
On the other hand, the actual turbulent motions (and the associated inertial range for the forward cascade of energy) are expected to be constrained to lengths bounded above by $L_O$, and so a reasonable estimate for $L_i$ should be $L_O$. 
Therefore,  (\ref{2Lbalance}) reduces to 
\begin{equation}\label{LH_ROT}
\Gamma_{\mathcal M} \propto R_{OT}^{-1}
\end{equation}
for this first `young turbulence' phase. 

\cpc{\subsection{`Fossilization' scaling for $\Gamma_{\cal M}$}\label{sec:foss}}
Conversely, during the final decaying or `fossilization' phase when $L_T  \ll L_O$, it seems reasonable that the mixing length can still be identified with the overturning scale, so  $L_m \sim L_T$. In this stage, $L_T$ can also be closely related to the integral scale of the turbulence, $L_i$, as the energy for turbulence and mixing is injected by the remaining, and smaller scale than earlier in the flow evolution, overturnings.  $L_O$ may now be interpreted as the largest eddies which could in principle be present within the given  ambient stratification.
However, for such eddies actually to arise, energy has to be injected in to the system at a sufficient rate, 
and since $L_T$ has dropped below $L_O$, there is no longer a mechanism by which that injection can occur. \cpc{Crucially, this should not be interpreted as `strong' stratification suppressing (in particular) vertical motions \cpcf{inducing strongly stratified anisotropic flow}, but rather that turbulent dissipation has converted a large proportion of the kinetic energy, and there is (at the moment) no forcing or energy injection mechanism occuring.}
Such a  gap between the actual eddy length scale (scaling with $L_T$) and the notionally possible eddy length scale (scaling with $L_O$, if only the flow was energetic enough) is one of the  hallmarks of `fossil turbulence'~\citep{gibson1987fossil}. It is important to remember that both $L_T$ and $L_O$ decrease with time, but $L_T$ decreases more rapidly, and so the  $L_O$ is `left behind' by the more rapid decay of the energy injection scale associated with $L_T$.

Thus, in the final fossilization phase, (\ref{2Lbalance}) reduces to
\begin{equation}\label{RH_ROT}
\Gamma_{\mathcal M} \propto R_{OT}^{-4/3},
\end{equation}
where once again it is important to remember that this argument is \cpc{fundamentally} associated with \cpcf{the  stratification always being below some critical, threshold strength},
not least because of the implicit assumption that buoyancy and momentum are mixed on the same scales and by the same processes.
In summary, we thus suppose that equation (\ref{LH_ROT}) holds in the $R_{OT} \ll 1$ limit, corresponding to the first growing `young turbulence phase', while (\ref{RH_ROT}) holds in the $R_{OT} \gg 1$ limit, corresponding to the final decaying or `fossilization phase', \cpc{with the variation in the dependence of the flux coefficient with 
$R_{OT}$ being associated with different (temporal) phases in the evolution of the  \cpcf{overturning} stratified mixing event, \cpcf{where buoyancy forces never dominate the dynamics}.}

\cpc{\subsection{`Goldilocks' scaling for $\Gamma_{\cal M}$}}
As previously reported in observations \citep[e.g. see][]{Dillon1982VerticalScales,ferron1998mixing}, \cpc{and further confirmed by our analyses of these oceanographic datasets,} most of the observed overturns actually exist around $R_{OT}\sim 1$,
which, from \S \ref{phenom}, occurs during the intermediate, energetically mixing phase. 
As pointed out in \citet{mashayek2017role}, this also actually turns out to be a regime of {\it efficient} (indeed apparently optimal) mixing, which we refer to as `Goldilocks' mixing. To achieve a parameterization for this regime, we simply combine (\ref{LH_ROT}) and (\ref{RH_ROT}) using the following analytic formula:
\begin{equation}\label{G_ROT}
\Gamma_{\mathcal M} =A\frac{R_{OT}^{-1}}{1+R_{OT}^{\frac{1}{3}}},
\end{equation}
which is constructed to have the correct asymptotic properties at small and large $R_{OT}$.

Before discussing the physical interpretation of (\ref{G_ROT}) and attempting to determine the  scaling coefficient `A' on physical grounds, it is very important to appreciate  that many of the arguments presented above are not new. In fact, employing $R_{OT}$ as a turbulence age proxy goes back several decades as reviewed by \citet{smyth2000length}, who also proposed that it be used to quantify $\Gamma_{\mathcal M}$. Numerical simulations and observations have also shown clear dependence of mixing on $R_{OT}$ \citep{smyth_etal_2001,mashayek2017role,Ijichi2018ObservedOcean,Ijichi2020HowAbyss,Smith2020AMixing}. Equation (\ref{RH_ROT}), in particular, has been previously proposed by many authors \citep{Weinstock1992VerticalKuo,Schumann1995TurbulentFlows,Baumert2000Second-momentFlows,Ijichi2018ObservedOcean}. 

\cpc{Indeed, these previous authors typically invoked turbulence properties largely unaffected by `weak' stratification. We go a step further by arguing that this particular scaling is appropriate only late in the life cycle of a time-varying overturning mixing event, \cpcf{where the stratification is only required not to be so strong that it always remains sub-dominant in the flow evolution}.}
Furthermore, 
$L_T$ was argued by \cite{Ivey1991650} as representative for the actual mixing length scales, (once again without explicit consideration of time dependence) an idea which was verified later via numerical simulations \citep[see][for example]{Shih2005}. However, we are not aware of   a previous proposal of equation (\ref{LH_ROT}) for `young', yet vigorously overturning mixing events, and as we shall show, its combination with (\ref{RH_ROT}) leads to an accurate parametric modelling of  $\Gamma_{\mathcal M}$  through (\ref{G_ROT}) (or indeed alternative definitions like $\Gamma_{\chi}$) in a way that would not be possible merely based on (\ref{RH_ROT}). 

\cpc{\subsection{Context in terms of previous studies}\label{sec:previous}}
It should also be noted that while we use the paradigm of a single three-phase mixing life cycle for arguments herein, our arguments also hold in a system composed of a multitude of interacting, co-existing overturns, \cpc{at different stages of their individual life cycle}. In such systems, the young phase simply refers to the phase where large quasi-laminar and relatively recently formed overturns exist and provide APE to small scale overturns to grow, while the fossilization phase corresponds to when the APE source has been largely depleted and the actual still-occurring overturns are smaller than the notionally possible size scale as marked by $L_O$. 

\citet{Mater2015BiasesData} showed, using several oceanic datasets, that the evolution of $R_{OT}$ in data does in fact resemble that based on the shear-driven KHI induced mixing flow evolution shown in figure  \ref{Fig1}. In a companion paper, \citet{Scotti2015BiasesSimulations} argued that in shear-driven mixing $L_T\sim L_O$, whereas in convectively-driven mixing (where the underlying source of TKE is APE), $L_T$ can actually overestimate mixing. \cpc{However, we would argue} such a clear \cpc{and binary} distinction between the two types of mixing is not entirely appropriate. A simple mixing event due to a shear instability can be dominated by a cascade of `secondary' shear instabilities growing on shear instabilities, for example as in an energetic estuary as shown in \cite{Geyer_etal_2010}, or it can be more appropriately characterized as being convectively-driven once the primary billow has rolled up, as observed  in the thermocline by \cite{Woods_68}. We argue that  if a distinction is allowed in principle between  the mixing length and the turbulent integral scale, as done here,
at least some `convective' mixing can fit within the same parameterization as `shear-driven' overturnings; oceanic datasets appear to support this argument. Needless to say, our argument does not extend to truly convection-driven mixing such as that in deep convection zones, but rather to patches of mixing where thinking of relatively isolated \cpc{time-varying} `overturning' mixing events \cpcf{where the stratification is sufficiently subcritical not to suppress such overturning} is reasonable.

\citet{Caldwell1983OceanicCreation} provided a nice description of the observed data on $R_{OT}$  on physical grounds and summarized the opposing views that became a source of controversy: on one hand, Gibson \citep{gibson1987fossil,Baker1987SamplingTurbulence} argued that most observations were what he referred to as fossilized turbulence, while \citet{gregg1987diapycnal} argued that they were of active turbulence. Caldwell proposed that the largest eddies (of scale $L_O$) feed on the original overturn (of scale $L_T$). He provided scaling arguments, observational support, and other references, to show that $L_O\sim L_T$ in this growth phase. This was also pointed out by \citet{itsweire1993turbulence}. This is consistent with the idealized picture presented in figure \ref{Fig1},
as well as with simulations of other non-shear-instability-like turbulent overturns \citep[e.g. see][]{chalamalla2015mixing,fritts2003layering}. Caldwell argued that that while the fossilization phase is also captured in the data, most of the observed overturns are in a state of `continuous production'.

This `continuous production' corresponds to the intermediate
energetically mixing phase identified in \S \ref{phenom}.  As the mixing during
this phase seems both to be most efficient and also to be 
associated with $L_O \sim L_T$, we refer to this phase
as being the `Goldilocks' mixing phase, where
the turbulent overturning is neither too `hot' ($L_T \gg L_O$) nor too `cold' ($L_T \ll L_O$) but `just right' ($L_O \sim L_T$).  As argued in \citet{mashayek2017role}, in this Goldilocks phase, the energy is most efficiently supplied to turbulence as the external driving  (associated with the primary KHI billow roll up) is at a scale which essentially matches the upper bound of an assumed inertial subrange largely unaffected by the stratification for which the \cpcf{(dynamic)} scalar mixing \cpcf{of buoyancy} is subordinate to the turbulence cascade. 

\cpc{\subsection{Quasi-equilibrium and Self-Organized Criticality}\label{sec:soc}}
It has been suggested \citep{sherman1978turbulence,Turner1973BuoyancyFluids,Salehipour2018,Smyth2019Self-organizedTurbulence,Smyth2020} that various kinds of shear-induced geophysically relevant turbulence \mha{are} in a state of  self-organized-criticality (SOC) in which the system is marginally unstable and overturns spontaneously emerge when the flow properties intermittently drop below a stability criterion with respect to a critical Richardson number, \cpc{i.e. in the sense discussed in the Introduction, the stratification intermittently becomes (just) subcritical, self-organizing and relaxing back towards criticality.}
Although \cite{Howland2018} demonstrated that 
KHI are exceptionally weak when the minimum 
Richardson number approaches the 
classical Miles-Howard \citep{Miles61,Howard61} criterion of $1/4$, calling into question the appropriateness of describing 
an energetically turbulent flow as being `marginally unstable' with respect to that specific value, as proposed by \cite{Thorpe2009}, there is a body of evidence 
that stratified turbulent flows can indeed adjust towards
a state where characteristic values of $Ri \simeq 0.17-0.25$
\citep{Zhou2017,vanReeuwijk2019,Portwood2019}. 
As originally argued by
\cpc{ \citet{Turner1973BuoyancyFluids} and 
\citet{sherman1978turbulence}}, this evidence is at least suggestive that shear-driven turbulent flows adjust to a `kind of equilibrium' with $Pr_T \sim O(1)$, and characteristic values of $Ri$ close to the 
linear stability critical value, perhaps fortuitously.

 We  combine these ideas to propose an appropriate value for the scaling coefficient `A'.
Keeping in mind that the $R_{OT} \ll 1$ and $R_{OT} \gg 1$ limits of equation (\ref{G_ROT}) represent the young and \cpc{decaying or} fossilizing limiting phases of a turbulent mixing event, $R_{OT}\sim 1$ thus represents the intermediate Goldilocks mixing phase. Let us assume that such a  mixing phase is 
in the `kind of equilibrium' proposed by Turner, which in the modern language of physics is essentially equivalent to a state of self-organized criticality.   
However, it might be described,  emergence of the balance $R_{OT}$ on physical grounds is consistent with the empirical
fact that the majority of observed overturns appear to be in this phase of efficient, and indeed `optimal' mixing, 
as  has been discussed previously~\citep{Dillon1982VerticalScales,ferron1998mixing,Mater2015BiasesData,mashayek2017role}. 

Therefore, we assume that  the flow has a characteristic critical Richardson number $Ri_{cr}$ close to, but perhaps somewhat less than  $1/4$, \cpcf{i.e. that the stratification of the flow is just slightly subcritical.} \cpc{Though it is perhaps stating the obvious, it important to appreciate that such a Richardson number should be considered as being significantly less than one, and to be associated with a flow where it is \cpc{not to be} expected that the turbulence 
\cpc{will be dominated}
by the stratification. For such a \cpcf{subcritically} stratified flow at} $R_{OT}\approx 1$, Eq. (\ref{G_ROT}) reduces to 
\begin{equation}
\Gamma_{gold}=\frac{A}{2}.
\end{equation}
Combining \eqref{eq:Kdef} and \eqref{eq:nutdef}, 
\begin{equation}
    Pr_T = \frac{P}{[-B]} \frac{N^2}{S^2} = \frac{Ri}{Ri_f}, \label{eq:rifdef}
\end{equation}
defining the `flux' Richardson number $Ri_f$ \citep{Turner1973BuoyancyFluids}.
Over sufficiently long averages and ignoring transport terms, we can assume $Ri_f$ is equivalent to a particular definition of 
a mixing efficiency, since $P \simeq {\mathcal M} +\epsilon$ and $-B \simeq {\mathcal M}$. Therefore 
\begin{equation}
    Ri_f \simeq \frac{\Gamma_{\mathcal M}}{1+\Gamma_{\mathcal M}} , \label{eq:rifdef2}
\end{equation}
\citep[see][]{Ivey1991650,Shih2005,Salehipour2019DiapycnalTurbulence}
Therefore,
\begin{equation}
    \Gamma_{\mathcal M} = \frac{\frac{Ri}{Pr_T}}{1- \frac{Ri}{Pr_T}}. 
\end{equation}
\cpc{Since we are always assuming that the 
\cpcf{mixing of scalar is slaved to the mixing of momentum}
so that $Pr_T\sim 1$, $Ri_f \sim Ri \lesssim 1/4$, and so $\Gamma_{\cal M} \lesssim 1/3$. As we shall see, the underlying
assumption that the flow is `\cpcf{subcritically} stratified' but time-evolving, leading to these various scalings, must always be remembered when comparing to previous studies, particularly those of \cite{Maffioli2016MixingTurbulence} and \cite{Garanaik2019OnFlows} based around the turbulent Froude number $Fr_T = {\epsilon}{/[N k]}$.}

If we further assume  that the `Goldilocks' optimally efficient mixing occurs at some specific $Ri_{cr}$ \cpc{rather than over some range bounded above by $Ri_{cr}$}, consistently
with the flow being in a self-organised critical equilibrium, we can then express the scaling parameter `A' as
\begin{equation}\label{A}
A=\frac{2\frac{Ri_{cr}}{Pr_t}}{1-\frac{Ri_{cr}}{Pr_t}}.
\end{equation}
Evidence from a range of flows suggests that vigorous turbulence
in a \cpcf{subcritically} stratified flow (i.e. \cpcf{in the specific sense that the flow has} sufficiently small $Ri$ that
the turbulence, characterized by overturnings, can be sustained throughout the flow)
has $Pr_T \simeq 1$ \citep{Zhou2017,vanReeuwijk2019,Portwood2019}. 
There is also evidence that appropriate values of $Ri_{cr}$
range from $0.16$ (as reported in \cite{Portwood2019}) through
$0.21$ \citep{Zhou2017,vanReeuwijk2019} to the \cpcf{canonical} `marginally stable'
value of $1/4$ \citep{Salehipour2018,Smyth2019,Smyth2020}.
Indeed, due to a variety of reasons, in particular associated with the presence
of ambient turbulence, the critical value of the Richardson number associated
with linear instability may be expected to be less than \cpcf{this} canonical
value of $1/4$ \citep{smyth_etal_2001,thorpe2013effect,Kaminski2019StratifiedTurbulence}.

Using these expressions $Pr_T \simeq 1$  and $1/6 \lesssim Ri_{cr} \lesssim 1/4$ suggests that $2/5 \lesssim A \lesssim 2/3$,
i.e. $1/5 \lesssim \Gamma_{\mathcal M} \lesssim 1/3$,
with the canonical lower value being associated with $Ri_{cr} =1/6$ and $Pr_T =1$, consistently
with the direct calculations of \cite{Portwood2019} in a continuously forced flow,
and also the (apparently) most efficient
mixing in an evolving Kelvin-Helmholtz unstable shear flow \citep{Mashayek2013Time-dependentFlux}. 
Alternatively $\Gamma_{\mathcal M}=0.2$ can also be identified with $Ri_{cr}=1/4$, $Pr_T=5/4$,
which is also largely consistent with available data, as one might argue that a continuously forced  flow should correspond to the intermediate mixing phase of an evolving overturning event.
\color{black} As we will soon show, for the data considered in this work, `A' lies within $2/5 \lesssim A \lesssim 2/3$ for individual and combined datasets. This implies that the above assumptions about $Pr_T$, $Ri_{cr}$ are reasonable.\color{black}

\cpc{At the risk of belabouring the point,} it is also important to appreciate \cpc{two fundamental aspects} of our argument concerning this optimal mixing phase with intermediate values of $R_{OT} \sim O(1)$ between the two asymptotic regimes of `young' turbulence and the decaying `fossilization' phase. \cpc{First, although} this 
regime is indeed intermediate, it still occurs when the flow 
\cpcf{has} relatively small characteristic Richardson numbers
\cpcf{significantly less than one}, with crucially $Pr_T \simeq 1$, \cpc{and so also $Ri_f \simeq Ri$.} Therefore, the mixing of buoyancy is set by the mixing of momentum,
\cpc{and the density is mixed essentially as a passive scalar, with buoyancy playing \cpcf{an at most sub-dominant} dynamical role. Specifically, and critically for our modelling}  the properties of the turbulence itself,
in particular characteristic length scales and time scales, are
largely unaffected by stratification in this intermediate  mixing phase within our framework. 
\cpc{Second, the time-dependence of the flow evolution is central to our argument, and this phase naturally follows the `young' (and crucially vigorously overturning) phase of the flow evolution, and precedes the later decaying or fossilizing phase. Fundamentally, at no stage in the flow evolution do buoyancy effects dominate \cpcf{\bf{any}} aspect of the flow evolution.}
\begin{figure}
\begin{center}
\includegraphics[trim=0 0 0 0,clip,width=.7\textwidth]{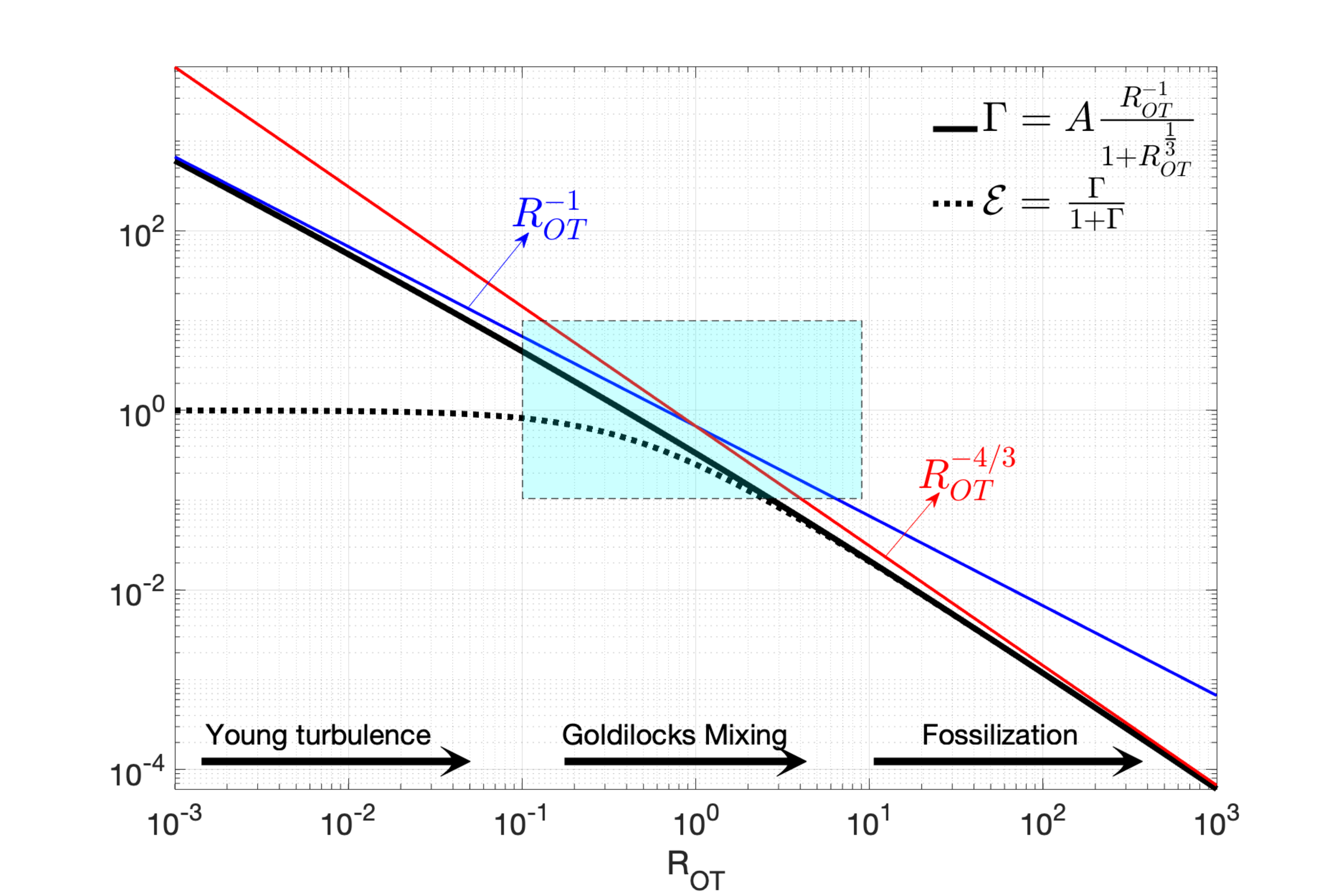} 
\caption{Variation with $\log{R_{OT}}$ of the parameterization (\ref{G_ROT}) (with $A=2/3$, implying $\Gamma_{\mathcal M}=1/3$ and $\mathcal{E}_i=1/4$ at $R_{OT}=1$). Also shown are the two asymptotic scalings for growing turbulence  (\ref{LH_ROT}) and for decaying turbulence  (\ref{RH_ROT}) with thin grey lines, and the corresponding mixing efficiency, with  a dashed red line. \color{black} The light blue box in the middle of the plot represents the bounding box within which the ocean turbulence data considered in this work lie, as will be shown in figure \ref{Fig3}.
\cpc{Conceptually, time for a particular shear-driven mixing event runs from left to right as $R_{OT}$ increases.}}
\label{Fig2}
\end{center}
\end{figure}

To illustrate our proposed parameterization,  figure \ref{Fig2} shows the parameterization in (\ref{G_ROT}) plotted for $A=2/3$ along with the 
asymptotic limits given by (\ref{LH_ROT}) and (\ref{RH_ROT}). Also shown is the naturally equivalent definition for an (instantaneous) mixing efficiency, i.e. \begin{equation}
\mathcal{E}_i \equiv \frac{\Gamma_{\mathcal M}}{1+\Gamma_{\mathcal M}} .\label{eq:egam}    
\end{equation}
This definition of mixing efficiency  tends to one in the limit of a non-turbulent flow due to non-zero molecular diffusion (at $Pr=1$) but the complete absence of turbulent dissipation, hence yielding a perfectly efficient (laminar) mixing, but importantly, {\bf no} turbulent mixing. \cpc{Yet again, it is important to remember that
the flows of interest \cpcf{should never be thought of as strongly} stratified, and so $R_{OT} \rightarrow 0$ because $\epsilon \rightarrow 0$ in the numerator of  the Ozmidov length $L_O$ as there is no turbulence at all, not because $N$ is very large in the denominator, thus suppressing the turbulence due to buoyancy forces.}
 In the limit of $R_{OT}\rightarrow \infty$, $\mathcal{E}_i$ tends to zero as would be expected in  decaying turbulence, \cpc{and $R_{OT}$ since the Thorpe length $L_T$ is decaying more rapidly that $L_O$. Finally, } at the heart of the intermediate `Goldilocks' phase, $R_{OT}=1$, $\mathcal{E}_i=A/(2+A)$.  

\begin{figure}
\begin{center}
\includegraphics[trim=0 0 0 0,clip,width=1.\textwidth]{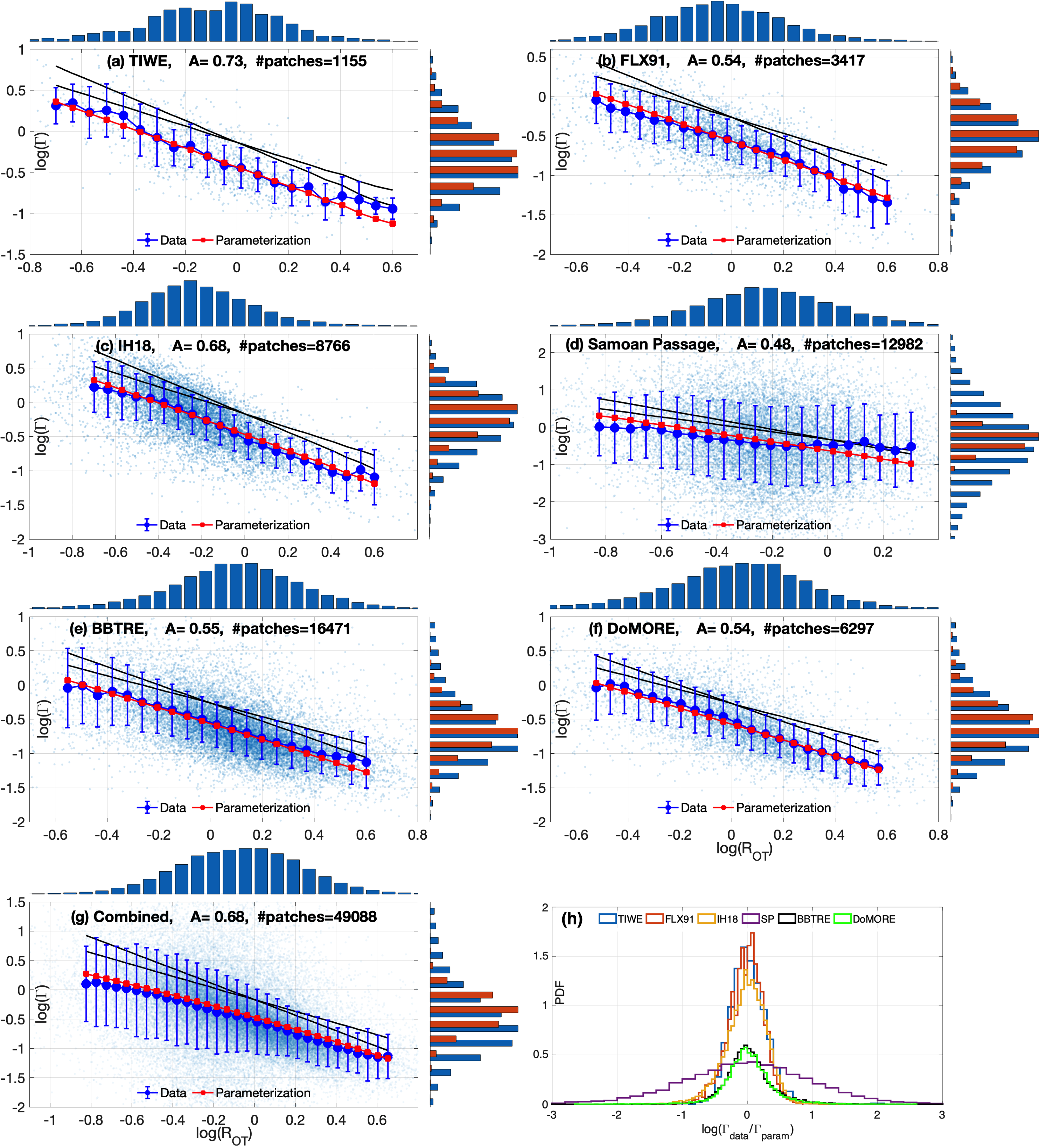}
\caption{Panels (a-g): Observationally inferred turbulent flux coefficient $\Gamma_{data}=\Gamma_\chi$ (constructed from 
measurements of $\chi$ and $\epsilon$)   as a function of $R_{OT}$ for six oceanic datasets  introduced in section \ref{sec:data}, as well as for the combined data. In each panel, blue scattered dots in the background represent actual patches, the solid blue line with filled circles and error bars represents the data binned (in log($\Gamma_\chi$) space), and the solid red line with square symbols represents the parameterized $\Gamma_{param}=\Gamma_{\mathcal M}$ using \eqref{G_ROT}. \color{black} Error bars represent $\pm 1$ standard deviation. The value of `A' for each panel is inferred based on linear regression of the corresponding data to log(Eq \ref{G_ROT})\color{black}. The right side inset compares PDFs of $\Gamma_{data}$ vs $\Gamma_{param}$   and the top inset shows PDF of data in log($R_{OT}$) space\color{black}. The solid straight black lines represent $\Gamma =A\ R_{OT}^{-1}$ and $\Gamma =A\ R_{OT}^{-4/3}$\color{black}. Panel (h) shows PDFs of $\Gamma_{data}/\Gamma_{param}$ for  the various datasets.}
\label{Fig3}
\end{center}
\end{figure}

\color{black}Figure \ref{Fig3}a-f shows an excellent agreement between the parameterized form of $\Gamma_{param}=\Gamma_{\mathcal M}$ using (\ref{G_ROT}) and measurements of $R_{OT}$ and the observationally inferred $\Gamma_{data}=\Gamma_\chi= \chi/\epsilon$ constructed from measurements of $\chi$ and $\epsilon$ `directly' from the six datasets introduced earlier. Each panel also includes the $\Gamma =A\ R_{OT}^{-1}$ and $\Gamma =A\ R_{OT}^{-4/3}$ limits. These two are expected to be relevant only in the limits of $R_{OT} \ll 1$ and $R_{OT}\gg 1$, as was shown in figure \ref{Fig2}. However, the datasets in figure \ref{Fig3} lie near the Goldilocks limit $R_{OT}\sim 1$ (the blue box in the middle of figure \ref{Fig2}). Thus, the two limiting cases  \cpc{are, perhaps unsurprisingly, inaccurate} in the range relevant to the data whereas their joint scaling, as formulated by Eq. \eqref{G_ROT}, \cpc{much better} represents the data. This is particularly clear when all the data are combined in panel (g) where a total of $\sim$50k turbulent patches provide  sufficient statistics to provide a fair comparison between the data and the parameterization (note that \cpc{at least a} few tens of thousands of samples are needed to capture the distribution statistically significantly, \cpc{as demonstrated by} \cite{CM21}).

As discussed earlier in connection with Eq. \eqref{A}, for $Pr_T \simeq 1$  and $1/6 \lesssim Ri_{cr} \lesssim 1/4$, we \cpc{obtain} $2/5 \lesssim A \lesssim 2/3$. As shown in the \cpc{captions} of individual panels in figure \ref{Fig3}, the range of values for `A' obtained based on regression of individual datasets to the parameterization \eqref{G_ROT} falls within this range except for TIWE which is data-limited from a statistical standpoint but nevertheless gives an `A' very close to the upper bound. The corresponding `A' for the combined data is almost exactly $2/3$ which corresponds to  $Pr_T \simeq 1$  and $Ri_{cr} \simeq 1/4$. Of course, this \cpc{close} agreement might \cpc{well} be somewhat fortuitous considering we only use six oceanic datasets. But given the wide range of processes, geographical locations, and oceanic depths spanned by these datasets, it is pleasing that their mixing can be captured so well using a parameterization entirely based on physical grounds: Eq. \eqref{G_ROT} depends on the Kolmogorov theory, the mixing length theory, and the turbulent kinetic energy budget, while `A' is determined based on a classical theoretical prediction for the critical Richardson number necessary for shear-induced \cpc{instability}. No empiricism, tuning, or filtering/modification of the data was required to obtain the agreement.\cpc{Indeed, allowing some empiricism associated with numerical simulations such as those reported inin \cite{Deusebio2015,Zhou2017,Portwood2019,vanReeuwijk2019} so that the critical Richardson number $Ri_{cr} \sim 0.2$ would nevertheless only modify the result slightly.}\color{black}

\cpc{Crucially, this agreement} also suggests that ocean mixing likely occurs at scales sufficiently small that the local gradient Richardson number falls below a critical value so that mixing in such a state remains close to \cpcf{and perhaps slightly below} marginally unstable, and \cpc{fundamentally}, scalings for the properties of the turbulence are 
\cpcf{never dominated}
by the ambient stratification.  The turbulence itself may be generated at a scale close to the shear instability, through for example mixing in the thermocline through the break-down of KHI. Alternatively, the initial generation  may be at a much larger scale through (for example) tidal generation, which through linear and nonlinear processes, leads to generation of shear at sufficiently small scales at which local $Ri$ can become sufficiently small \citep{Nik_Legg_GRL_10}. In other words, we hypothesize that the classic paradigm of a parallel shear flow, as idealized as it undoubtedly is, may plausibly lie at the heart of ocean interior mixing. 

To compare the data and parameterization \cpc{more quantitatively}, figure \ref{Fig3}h shows the histograms of \mha{the} ratio of $\Gamma$ from data and the parameterized $\Gamma$. The agreement is remarkable given the range of dominant driving mechanisms for the different datasets: shear instability triggered by tropical instability waves in the equatorial undercurrent (TIWE), near-inertial-shear  (FLX91), internal tides (BBTRE), and hydraulically controlled flow over abyssal canyons and sills (DoMORE and Samoan Passage). For the latter two cases, the agreement is somewhat surprising considering that the measured turbulence is close to the seafloor and possibly partially convectively driven, although even such convectively breaking overturns are shear-driven in the sense that the background shear is a main source of energy. Nevertheless, it is reasonable to expect an influence of another length scale, which is absent in the construction of (\ref{G_ROT}) for near boundary turbulence. Perhaps this explains the larger spread in the data for the Samoan Passage than predicted by  (\ref{G_ROT}). Nevertheless, the agreement remains good at the peak of the PDF of the data where $R_{OT}\sim 1$.

\subsection{Comparison to $Fr_T$-based Parameterizations}
\color{black}
\cpc{ Here, we have been focussed on mixing events which can 
be related to shear-driven processes 
\cpcf{with subcritical stratification}
in the \cpcf{specific} sense that vigorous over-turning events can be observed. As cogently argued by \cite{Maffioli2016MixingTurbulence}, building indeed on the insights of \cite{Ivey1991650}, there are several reasons why it is   appropriate
to attempt to construct  parameterizations in terms of 
 the turbulent Froude number $Fr_T$ defined
(using our notation) as 
\begin{equation}
    Fr_T \equiv \frac{\epsilon}{N k}. \label{eq:frtdef}
\end{equation} 
They observed (from the results of body-forced numerical simulations) that $\Gamma \propto Fr_T^{-2}$ for large $Fr_T$, 
consistently also with  density current observations as discussed in \cite{Wells_2010}.}

\cpc{Of course, large $Fr_T$ corresponds to `weak' stratification, and so it is entirely expected that the unstratified scaling (\ref{intScale}) applies. Therefore, remembering
also the definition of $L_O$ (\ref{eq:lengthsdef}), appropriately large $Fr_T$ is proportional to
\begin{equation}
    Fr_T \propto \left ( \frac{L_O}{L_i} \right) ^{2/3} .\label{eq:fr1}
\end{equation}
}

\cpc{
Considering this weakly stratified class of flows, \cite{Maffioli2016MixingTurbulence} (effectively) assumed that $Pr_T \sim 1$, $\Gamma_\chi = \Gamma_{\cal M}$ and that the mixing length $L_m \sim L_i$ (using our notation), so that 
(\ref{eq:lengthsdef}) becomes
\begin{equation}
    \Gamma_{\cal M} \propto \left ( \frac{L_i}{L_O} \right) ^{4/3} \propto Fr_T^{-2} . \label{eq:maf}
\end{equation}
}

\cpc{
\cite{Garanaik2019OnFlows} extended the discussion of \cite{Maffioli2016MixingTurbulence}, constructing and interpreting various Froude-number scalings in terms of time scales and length scales, motivated by  constructing a parameterization that would be practically useful in terms of oceanic measurements. In the high $Fr_T$ limit, they also argued (in our notation) that $L_i \sim L_T$, and so that $\Gamma_{\cal M} \propto Fr_T^{-2} \propto R_{OT}^{-4/3}$. 
This corresponds to the final decaying fossilization phase described above, though neither \cite{Maffioli2016MixingTurbulence} nor
\cite{Garanaik2019OnFlows} considered time-dependence of mixing events (by design).
In constrast to our inherently time-dependent viewpoint, their argument was that $L_T < L_O$ was a generic characteristic of weakly stratified turbulent flows, whereas we postulate that this scaling  only occurs late in the flow evolution of a shear-driven overturning mixing event, 
\cpcf{and in point of fact, our arguments only require
the stratification to be subcritical (and thus allow shear-driven overturnings) with sufficiently small but  necessarily insignificant characteristic Richardson number.}
}

\cpc{
Furthermore, although not couched in terms of shear flows by \cite{Maffioli2016MixingTurbulence}, the $Fr_T^{-2}$ scaling can also be interpreted in terms of (\cpcf{sufficiently} small) Richardson numbers, as the natural equivalent scaling is that
\begin{equation}
    Fr_T \sim \frac{U_0}{L_i N} \propto Ri^{-1/2} ,\label{eq:frri}
\end{equation}
and so 
\begin{equation}
 \Gamma_{\cal M} \propto  Fr_T^{-2} \propto Ri . \label{eq:gamri}  
\end{equation}
This is of course entirely as expected for 
flows with $Pr_T \sim 1$, as then $\Gamma_{\cal M} \lesssim  Ri_f \simeq Ri \lesssim 1/4$.
}

\cpc{
 Interestingly, \cite{Garanaik2019OnFlows} also presented scaling arguments for two other (effectively assumed to be quasi-steady) classes of flows, neither of which \cpcf{should be construed as being equivalent} to the 
(inherently time-dependent) phases of a \cpcf{subcritically} stratified flow \cpcf{as} considered here. Building on the observations of \cite{Maffioli2016MixingTurbulence}, which in turn were motivated by 
experimental observations dating back (at least) to \cite{Kato1969},
(see also \cite{park1994turbulent,Oglethorpe_2013,olsthoorn2015vortex})
\cite{Garanaik2019OnFlows} presented a scaling argument for the observed constant flux coefficient $\Gamma_{\chi} \propto Fr_T^{0}$ in (at least apparently) strongly stratified flows as $Fr_T \rightarrow 0$.  They argued that the strong stratification completely dominates the dynamics, so that both the turbulent dissipation rate $\epsilon$ and the buoyancy flux $B$ (and hence $\chi$ as the flow is quasi-steady) scale with $w^2 N$, where $w$ is the (perturbation) vertical velocity. Therefore $\Gamma_\chi$ is constant in very strong stratification. Clearly, this class of flows is not considered accessible by the (vertical) shear-driven mixing events
with \cpcf{subcritical stratification allowing overturnings to reach significant amplitude}  on which we are focusing in this paper.
}

\cpc{
Furthermore, as argued in more detail in \cite{Caulfield2021LayeringFlows}, care must be taken in inferring the existence of this class of strongly stratified and yet vigorously turbulent flows. The motivating experimental 
data were typically extracted from flows with  emergent density staircases, where relatively well-mixed and deep `layers' are separated by relatively thin `interfaces' of enhanced density gradient.  Indeed, as demonstrated by \cite{Portwood2016}, flows with low $Fr_T$ (when averaged over relatively large scales) typically exhibit significant spatio-temporal variability, with vigorous turbulence (characterised by large local values of $\epsilon$) being
restricted to patches of locally \cpcf{significantly weaker} stratification. 
}

\cpc{
Connecting this (assumed) strongly stratified class of flows
as $Fr_T \rightarrow 0$ and the weakly stratified class
with $Fr_T \gg 1$, \cite{Garanaik2019OnFlows} argued for the existence of an intermediate `moderately stratified' class of flows, where the properties are effectively hybrid between the two end-members. They argued that the buoyancy variance destruction rate should scale as in the strongly stratified limit so that $\chi \propto w^2 N$, while the turbulent kinetic energy (and its dissipation rate) are both (largely) unaffected by the stratification, so $k \sim w^2$, and
so, in this class of flows, they argued that
\begin{equation}
    \Gamma_\chi \propto \frac{kN}{\epsilon} = Fr_T^{-1} .\label{eq:intermediate}
\end{equation}
Furthermore, they  argued that this `moderately stratified' class is also
characterised by $Fr_T \sim O(1)$, which finally allowed them to argue that this intermediate class should have the length scaling
$\Gamma_\chi \propto R_{OT}^{-1}$ as when $Fr_T \sim O(1)$ it is expected that $R_{OT} \sim O(1)$ as well. 
}

\cpc{
The presented evidence of this intermediate class is perhaps not as convincing as for the two end members. Whatever the quality of the data fit, it is crucially important to appreciate that the arguments leading to this scaling are fundamentally different from our arguments leading to the (superficially same) scaling (\ref{LH_ROT}) for the early-time `young' phase of an inherently time-dependent and 
\cpcf{subcritically} stratified shear-driven mixing event. 
Furthermore, our \cpcf{`subcritically stratified'}
intermediate `Goldilocks' phase is also in marked conceptual contrast to the intermediate, `moderately stratified' class of flows parameterized by   
\cite{Garanaik2019OnFlows}, in the sense that the physical processes and balances should be thought of as arising for fundamentally different reasons. 

In their argument,  the stratification (and hence the characteristic values of the buoyancy frequency) in their intermediate class has an order one effect on the mixing.
\cpcf{That can be seen perhaps most clearly in terms of the implications of their scaling arguments for  $Pr_T$.
Since they  argue that $\chi \propto w^2 N$,
\begin{equation}
    \kappa_T \propto \frac{\chi}{N^2} \propto \frac{w^2}{N }.\label{eq:gvkt}
\end{equation}
On the other hand, since they argue that the turbulent kinetic energy $k \propto w^2$ is largely unaffected by stratification, it is natural
to suppose that the mixing length for momentum is given by
(\ref{intScale}), and so
\begin{equation}
    \nu_T \propto \left (\frac{w^3}{\epsilon} \right )  w \rightarrow Pr_T \propto \frac{N w^2}{\epsilon} \propto Fr_T^{-1} ,\label{eq:gvnut}
\end{equation}
using the scaling (\ref{eq:gvkt}) for $\kappa_T$.

It could perhaps  be argued that since in this regime $Fr_T \sim 1$, $Pr_T \sim 1$ still using this scaling. Nevertheless,
a consistent interpretation of this `moderately stratified' class of flows leads to the prediction that the dynamics
of the mixing of the buoyancy (as described by $Pr_T$)
does depend on the stratification, with in particular
$Pr_T$ increasing as the stratification becomes stronger, and hence $Fr_T$ decreases. Importantly such a variation in $Fr_T$ can occur straightforwardly for different quasi-steady forced flows within their assumed class of flows.

This is a qualitatively
different mixing behaviour from the mixing behaviour we argue occurs
during the intermediate (in time) `Goldilocks' mixing phase. Our argument implicitly assumes that the stratification is sufficiently subcritical that $Pr_T \simeq 1$ throughout the mixing life cycle, and in particular during the intermediate `Goldilocks' phase. During this phase,  it appears that the stratification 
is in some sense self-organized close to  a critical 
value of the Richardson number, and so there is not expected to be any parameter 
dependence of $Pr_T$: the flow is after all neither too hot nor too cold, but just right. 
In summary, their argument assumes an intermediate quasi-steady class of flows with hybrid dynamical balance between buoyancy and turbulence, while our intermediate (in time) phase is still (and always) subcritically stratified. We argue that it is the middle phase of an inherently time-dependent flow evolution, that happens after a primary shear-driven overturning breaks down, but before it enters a final fossilization decay phase.}

Indeed, 
caution must always be shown in drawing any connections between descriptions in terms of $Fr_T$ and $Ri$, particularly as $Ri \simeq 1$ is actually really a quite strongly stratified
flow in the context of shear-driven mixing events. Of course, there is no necessity for equality in the various scalings used in the construction of the parameterizations, and the presence of an order one constant can always allow $Ri$ to be sufficiently small \cpcf{to be subcritical in the sense used in this paper, i.e.} for a shear-driven overturning to develop \cpcf{to sufficiently large amplitude to undergo vigorous turbulent breakdown}, which we believe is actually the dominant process far from boundaries. }

In summary, while equation \eqref{RH_ROT} was proposed previously, to our knowledge equations (\ref{2Lbalance}, \ref{LH_ROT}), and thereby \eqref{G_ROT} are new. Furthermore, the previously proposed scaling relation only strictly applies for $R_{OT}\gg 1$ which corresponds to a small fraction of the data. Through proposing a scaling for the $R_{OT}\ll 1$ limit and combining the two limits,  we have managed to propose a comprehensive parameterization and also to determine the coefficient of proportionality on physical grounds. Note that earlier works used \eqref{RH_ROT} with an adjustable coefficient that could be tuned to fit the data better.
Figures \ref{Fig2} and \ref{Fig3} suggest that such a tuning is incorrect, and the data `feels' both (\ref{LH_ROT}) and  (\ref{RH_ROT}). In the next section, we will further discuss the significance of \eqref{LH_ROT} for bulk estimates of mixing on regional and global scales.
\color{black}



\color{black}
\section{Three key points}
\label{3points}
\color{black}

\subsection{The devil is in the tails}
An important question, one mostly left out of the literature discussing properties of various definitions of the turbulent flux coefficient $\Gamma$,  is the extent to which the nuances of variations in $\Gamma$ with other parameters matter in practice. There is no simple answer to this question, as it depends on the application for which the rate of mixing is being measured. Furthermore, a key missing piece is how one would connect the fluid-dynamical understanding of mixing of turbulent patches (as studied herein and in articles cited throughout), to an appropriate time and space averaged rate of mixing. The latter is out of the scope of our paper (but some discussion is provided in \citet{CM21}). Since our primary motivation behind this work is to relate efficiency of mixing to transformation of water masses in the ocean interior, which depends on the vertical divergence of the buoyancy flux \citep[see][]{MFNP_2015,Ferrari2016turning}, we attempt to answer the question by simply quantifying the extent to which various measures of $\Gamma$ will affect the total buoyancy flux obtained by summing the flux due to all the overturns in the datasets that we consider here. 
\begin{figure}
\begin{center}
\includegraphics[trim=0 0 0 0,clip,width=1.0\textwidth]{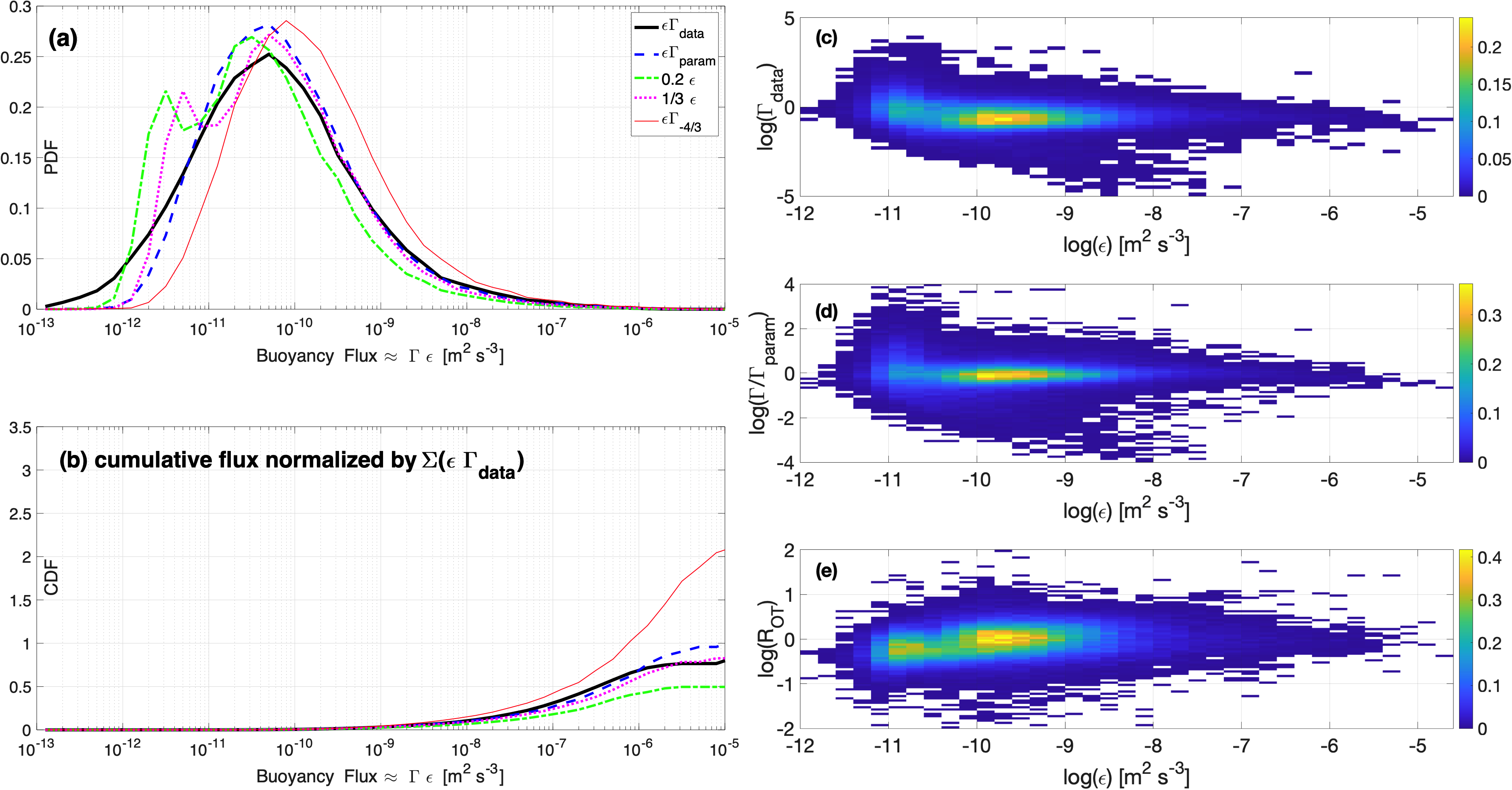}
\caption{(a) $\Gamma \times \epsilon$, considered a proxy for the buoyancy flux through the Osborn relation (\ref{eq:prdef}), calculated based on $\Gamma_{data}$, $\Gamma_{param}$ using our full parameterization (\ref{G_ROT}) and $\Gamma_{-4/3}$ using only the asymptotic scaling  (\ref{RH_ROT}), as well as based on use of constant values $\Gamma=0.2$ and $\Gamma =1/3$. The last choice is based on equation (\ref{A}) if we consider $Ri_{cr}=1/4$. (b) Cumulative flux for the cases of panel (a), all normalized by the total $\Gamma_{data} \epsilon_{data}$  both  obtained directly from data. (c,d,e) show bivariate histogram plots for the data over $\Gamma-\epsilon$, $\Gamma_{data}/\Gamma_{param}-\epsilon$, and $R_{OT}-\epsilon$ parameter spaces. All plots are made for the data combined for all six experiments in Table 1.
}
\label{Fig4}
\end{center}
\end{figure}
\begin{table}
\centering
\begin{tabular}{ l|c|c|c|c} 
  & $\Sigma(\epsilon\Gamma_{param})/\Sigma(\epsilon\Gamma_{data})$ & $\Sigma(0.2 \epsilon)/\Sigma(\epsilon\Gamma_{data})$ & $\Sigma(\frac{1}{3}\epsilon)/\Sigma(\epsilon\Gamma_{data})$ & $\Sigma(\epsilon\Gamma_{-4/3})/\Sigma(\epsilon\Gamma_{data})$  \\  
 TIWE & 0.96 & 0.53 & 0.89 & 2.01 \\ 
 FLX91 & 0.97 & 0.72 & 1.16 & 2.02  \\
 IH18 & 1.03 & 0.41 & 0.67 & 2.62  \\
 Samoan Passage & 2.91 & 1.21 & 1.99 & 6.58  \\
 BBTRE & 0.96 & 0.52 & 0.86 & 2.96   \\
 DoMORE & 0.91 & 0.24 & 0.39 & 2.27  \\
 Combined {\tiny} &  1.43 & 0.63 & 1.05 & 3.2  \\
\end{tabular}
\caption{Comparison of the inaccuracies between the total buoyancy fluxes using the Osborn approximation $\Gamma \epsilon$ estimated based on constant values and various other parameterizations of $\Gamma$.}
\label{Flux_c}
\end{table}

Taking $\Gamma \epsilon$ (for various definitions of $\Gamma$) as a proxy for the buoyancy flux $\kappa N^2$ through the Osborn balance (\ref{eq:prdef}), in figure \ref{Fig4}a we plot the flux based on  $\Gamma_{data}=\Gamma_{\chi}$ from the data, parameterizations, and as constants. (While we of course appreciate the many questionable assumptions underlying the approximation of actual  flux with $\Gamma \epsilon$, as discussed  in \citet{MP13,MCP13} for example, they do not affect our arguments here in any significant way.) As expected based on figure \ref{Fig3}, flux parameterized using (\ref{G_ROT}) agrees well with the observed flux, while flux parameterized using (\ref{RH_ROT}) is biased high. Most importantly, if we consider the total flux due to all overturns, given the several orders of magnitude that $\epsilon$ spans, the tail of the histograms makes a significant contribution and so the extent to which $\Gamma$ is accurately estimated for the tails is key. As shown in panel (b), it is the higher $\epsilon$ tail that leads to an overestimate when using parameterization (\ref{RH_ROT}). It is also notable that, while $\Gamma=0.2$ underestimates the total flux, $\Gamma=1/3$ actually leads to a much better agreement. Since $\Gamma=1/3$ is linked (at least within our modelling approach) to assuming that the critical Richardson  number  $Ri_{cr} \simeq 1/4$, as was discussed above, it seems a more appropriate choice in the simplest approach of using a single
constant value of $\Gamma$. 

Figure \ref{Fig4} includes 5 of the datasets combined. 
The  disagreement between various estimates, however, can be much larger when looked at case by case. To do so, in Table \ref{Flux_c} we compare the total flux based on various estimates (i.e. the high end of the curves in figure \ref{Fig4}b) for individual datasets. It is clear that parameterization (\ref{G_ROT}) is quite accurate in capturing the total flux for all cases (within 10$\%$). Even for the Samoan Passage case, which is somewhat of an outlier, it is still within a factor of 3. On the other hand, parameterization (\ref{RH_ROT}) is less accurate. $\Gamma=0.2$ underestimates the total flux by as much as a factor of four,  and fixing $\Gamma=1/3$ does a much better job. 

It is useful to interpret figure \ref{Fig4}a-b and the agreement  of data with parameterization (\ref{G_ROT}) by looking at the distribution of data over various parameter spaces, as shown in panels (c-e) of the figure. While the data are spread around $R_{OT}\sim 1$, the spread is much larger at low $\epsilon$, representing both the early stages of turbulence growth and the later stages of the turbulence after significant decay, as one would expect from consideration of figure 1. For energetic patches that correspond to the optimal `Goldilocks' mixing phase, $R_{OT}$ nicely collapses around the value $R_{OT} \simeq 1$  as shown in panel (e). For this Goldilocks phase, $\Gamma \sim 1/3$ as shown in panel (c) and in agreement with the assumption of $Ri_{cr}\sim 1/4$, while at early and later stages of turbulence life cycles, $\Gamma$ can be very large or very small, respectively. Panel (d) shows that the parameterization (\ref{G_ROT}) manages to collapse this spread on $R_{OT}$, and that importantly it works well for the tail of the histograms which make the largest contribution to the total flux. In summary, given the accuracy, simplicity, and physically-justified nature of the parameterization (\ref{G_ROT}), it seems like a natural way to model the turbulent flux coefficient $\Gamma$ associated  with overturns in shear-induced turbulence.

\color{black}
\subsection{Bulk $\Gamma$: the importance of young turbulence events}
\begin{figure}
\begin{center}
\includegraphics[width=1.\textwidth]{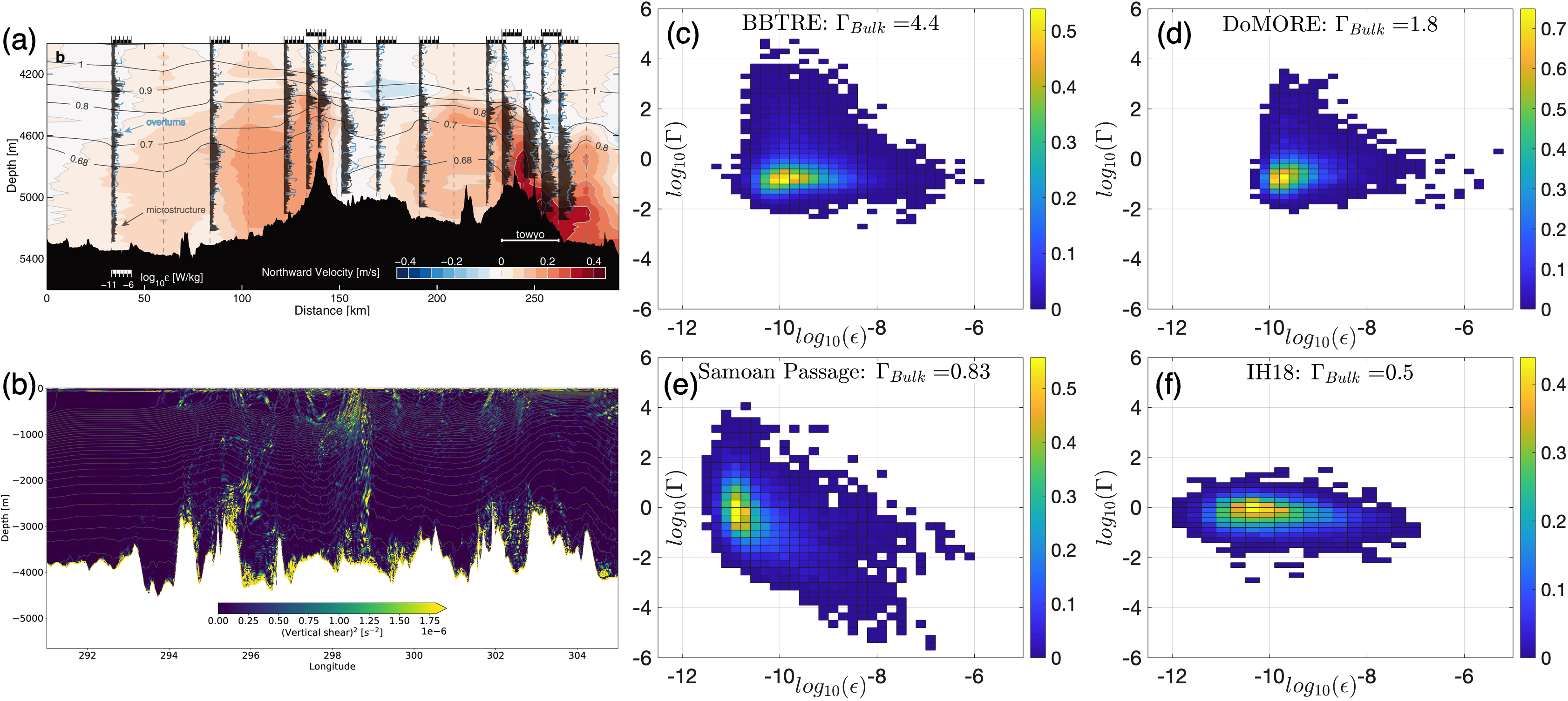}
\caption{\color{black}(a) An example of an energetic oceanic turbulence zone, in the Samoan Passage, to demonstrate the intermittency of turbulence; reproduced from \citet{alford2013turbulent}:
Northward velocity (colors), potential temperature (black contours), and dissipation rate measured by the velocity microstrucure profiles (shaded black profiles) and from Thorpe scales (blue profiles). The scale for $\epsilon$ is given at the lower left and redrawn above each profile. (b) A snapshot of vertical shear from an observationally forced and tuned wave-resolving simulation of the Drake Passage, borrowed from the work of \citet{mashayek2017topographic}. (c-f) Bivariate histograms showing the distribution of patches over the $\epsilon-\Gamma$ parameter space for four of the datasets considered herein. The values of bulk $\Gamma$ in each panel's title is obtained from Eq. (\ref{G_B}). Note that the data in panel e is associated with panel a.}
\label{Fig5}
\end{center}
\end{figure}

In ocean and climate general circulation models, small scale wave breaking is not resolved but the combined power available for mixing from tides, winds and other sources is available as a bulk value per grid cell. For example, an energetic turbulent zone in the ocean, such as that shown in figure \ref{Fig5}a for the Samoan Passage, is roughly represented by 1-2 grid cells in a typical climate model. A bulk $\Gamma$ is thus required to divide the power available (from tides, winds, currents, etc.) into bulk measures of mixing and dissipation for a given grid cell~\citep[see][for a through discussion]{Cimoli2019SensitivityEfficiency}. In empirical practice, this $\Gamma_{Bulk}$ is often taken to be $0.2$, 
even without all the variations and caveats associated with $\Gamma_i$ for individual turbulent patches, $\Gamma_{Bulk}$ inevitably depends on the statistics of turbulent patches within each grid cell. Several studies have applied parameterizations for $\Gamma$, derived based on patch data, to coarse grid-based calculations to establish the leading order importance of variations in $\Gamma$ for the larger scale ocean circulation~\citep{de2016impact,mashayek2017efficiency,Cimoli2019SensitivityEfficiency}. That approach, while illuminating,  is also not correct since it applies a patch-based quantity to grid cells of $\mathcal{O}(100\mathrm{km})\times \mathcal{O}(100\mathrm{km}) \times \mathcal{O}(100\mathrm{m})$ in size. In this subsection we show how $\Gamma_{Bulk}$ depends on statistics of turbulent patches and that  not only does it depend on Goldilocks mixing of energetic patches, but equally on the young patches which posses large $\Gamma$ and decaying patches with vanishing $\Gamma$. 

Imagining a typical energetic zone, such as the example in figure \ref{Fig5}a, we assume the domain to contain $n$ a number of turbulent patches (within a grid cell of a coarse resolution model which would represent the region) and over a period associated with a coarse resolution model time stepping. Panel (b) shows a snapshot from a wave-resolving high resolution realistic simulation to \cpc{demonstrate} the intermittency of vertical shear generated by surface winds and flow-topography interaction in the very energetic Drake Passage in the Southern Ocean where high flow speeds occur due to the constriction of the Antarctic Circumpolar Currents (ACC) and the overlying strong westerly winds known as the roaring forties. Even in such an energetic region, the flow is mostly non-turbulent (from a small scale stratified turbulence perspective-- a rich mesoscale and submesoscale turbulence field exists throughout the domain). However, in the `quiet' regions, background weak mixing \mha{almost always} exists: $97\%$ of the data acquired from $\sim$750 full-depth microstructure profiles from 14 field experiments considered in \citet{CM21} possessed a diffusivity larger than $10^{-6}$ $m^2/s$~\cite[also see][]{waterhouse2014global}. In the limit of completely laminar flow, $\kappa\rightarrow \mathcal{O}(10^{-7})\ m^2/s$ and so the flux $\mathcal{M}\approx\kappa N^2$ stays finite while $\epsilon$ tends to zero. Thus, while for energetic turbulent patches $\Gamma\approx \Gamma_{gold}$, for quieter regions it tends to $\mathcal{O}(10)$ and larger values (see histograms in figure \ref{Fig3}).

To show how the energetic, young and decaying turbulent patches collectively set the bulk $\Gamma$, we define
\be
\Gamma_{Bulk}=\frac{\mathcal{M}_{tot}}{\epsilon_{tot}}\approx \frac{\sum\limits_{i=1}^n \Gamma_i \times \epsilon_i}{\sum\limits_{i=1}^n \epsilon_i}
\label{G_B}
\ee
where $n$ represents the number of patches in a region of interest. Panels (c-f) of figure \ref{Fig5} show the estimated $\Gamma_{Bulk}$ for four deep energetic turbulent datasets considered herein. As discussed in figure \ref{Fig4}, for energetic patches with high $\epsilon$, $\Gamma_i \rightarrow \Gamma_{gold}$ while small $\epsilon$ corresponds to young and decaying patches which possess large and small $\Gamma_i$, respectively. For BBTRE and DoMORE, abundance of young patches result in a large, $\Gamma_{Bulk}$ even though most of the data corresponds to Goldilocks mixing. For IH18, a larger fraction of the data corresponds to Goldilocks mixing and so does $\Gamma_{Bulk}$. A simple thought experiment can help illustrate the impact of young turbulent patches. Imagine a box with three turbulent patches: a young patch ($\epsilon=0.001$, $\Gamma=100$), \cpc{a} moderately turbulent patch ($\epsilon=0.1$, $\Gamma=10$), and 
\cpc{an} energetic patch ($\epsilon=1$, $\Gamma=1/3$). This results in $\Gamma_{bulk}=0.44$.

Our key message is that while most of the collective community efforts have focused on finding the right parameterization for energetic turbulent patches, the net mixing also critically depends on young patches and the large fraction of the water column where only weak background turbulence \cpcf{occurs}. This highlights the importance of the new scalings presented in this work in \cpc{equations} (\ref{2Lbalance},\ref{LH_ROT}). It also highlights the need for investing efforts in  understanding \cpc{better} the statistics of turbulence to quantify its intermittency. Such statistics are key to connecting our understanding of physics of small scale stratified turbulence, as discussed in this work and the ones cited herein, to the larger scale applications. This statistical bridge is arguably highly underdeveloped- see \cite{CM21} for a discussion. Finally, while we used the patch data for four experiments in figure \ref{Fig5}, it was merely to illustrate the simple point that $\Gamma_{Bulk}$ can be large even though most patches have $\Gamma\sim \Gamma_{gold}$. We have no reason to believe that the number of patches within each experiment was statistically sufficient to quantify mixing \cpc{accurately} in the geographical region of each experiment. Also, the data we used are only for the already identified patches. The actual microstructure profiles also include regions that are not identified as patches, but contribute to $\Gamma_{Bulk}$.

\subsection{Statistics of $L_O/L_T$: active vs fossil turbulence}
Thus far we have repeated on several occasions and on empirical grounds (based on the data used in this study and other works cited herein) that `most of the data is centred around $R_{OT}\sim 1$'. This, however, lies at the heart of a historically significant debate. As discussed in the introduction, 
\citet{gibson1987fossil} and \citet{Baker1987SamplingTurbulence} argued that most observations were of fossilized turbulence, while \citet{gregg1987diapycnal} argued that they were of active turbulence. The latter,  referred to as `continuous production' by \citet{Caldwell1983OceanicCreation} and as Goldilocks mixing herein, is supported by a seemingly universal concentration of data around $R_{OT}\sim 1$. 

\begin{figure}
\begin{center}
\includegraphics[width=1.\textwidth]{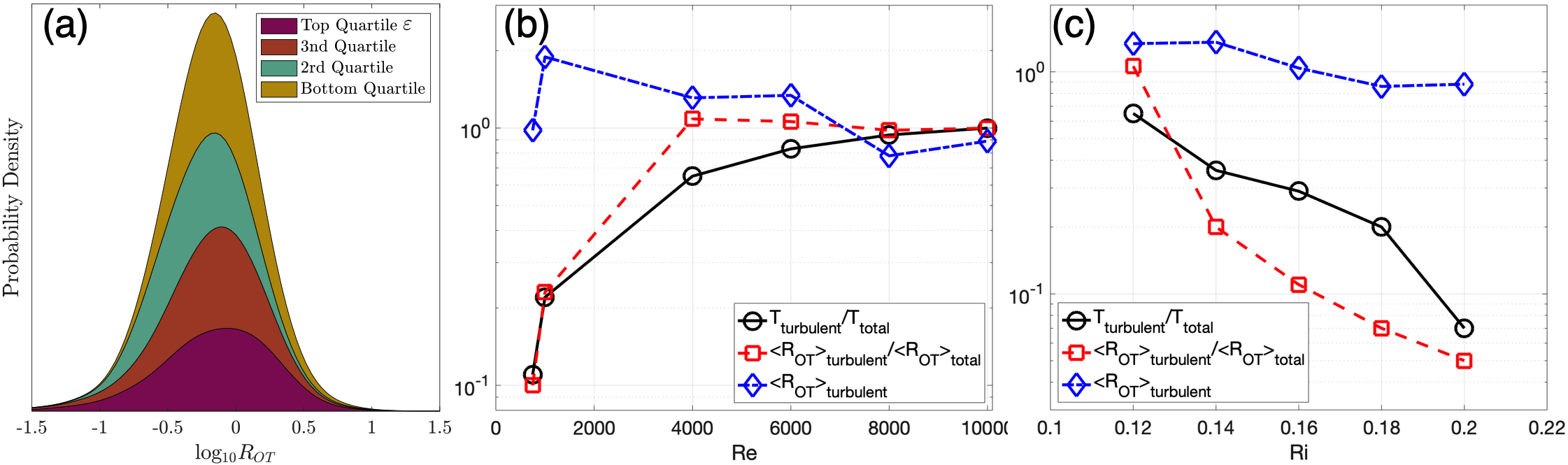}
\caption{\color{black}(a) Probability density function of $R_{OT}$, grouped into quartiles of $\epsilon$, from the combined six oceanic datasets considered in this work. (b,c) Temporal fraction of turbulence lifecyle as well as the ratio of $R_{OT}$ during turbulent phase of the flow to its mean value over the whole lifecycle. Each symbol represents a lifecycle-averaged quantity from a direct numerical simulation for a given pair of Reynolds and Richardson numbers. All cases in panel (b) are for $Ri=0.12$ while all cases in panel (c) are for $Re=6000$. Turbulent phase of the lifecycles are defined as the times when $Re_b>20$. Panels c,d are produced from analysis of simulations originally discussed in \cite{MP_POF,MP_GRL,MP13,MCP13}.}
\label{Fig6}
\end{center}
\end{figure}

Figure \ref{Fig6}a shows the histogram of the combined data considered in this work to highlight the clustering around $R_{OT}\sim 1$. Importantly, the clustering improves for more energetic turbulent patches with larger $\epsilon$. Panels b,c of the figure demonstrate, based on data from direct numerical simulations of shear instabilities, that once the parameter ranges approach those of oceanic turbulent patches (i.e. large $Re$ and subcritical $Ri$), a larger fraction of the total overturning lifecycle corresponds to sustained energetic turbulence (for which $R_{OT} \sim 1$). This explains the statistical distribution of $R_{OT}$ in panel a. In contrast, less energetic turbulence events, such as those created in laboratory or numerical simulations, or those observed in lakes, are expected to have a more broadly distributed (over $R_{OT}$) distributions. \cpc{Once again, we argue that $R_{OT} \sim 1$ is associated with an intermediate phase of an energetic \cpcf{time-dependent} \cpcf{subcritically} stratified mixing event, not an intermediate, hybrid class of `moderately stratified' \cpcf{quasi-steady} mixing events as argued by \cite{Garanaik2019OnFlows}.}

\section{A note on $\Gamma$ as a function of $Re_b$}\label{sec:Reb}
\begin{figure}
\begin{center}
\includegraphics[width=.75\textwidth]{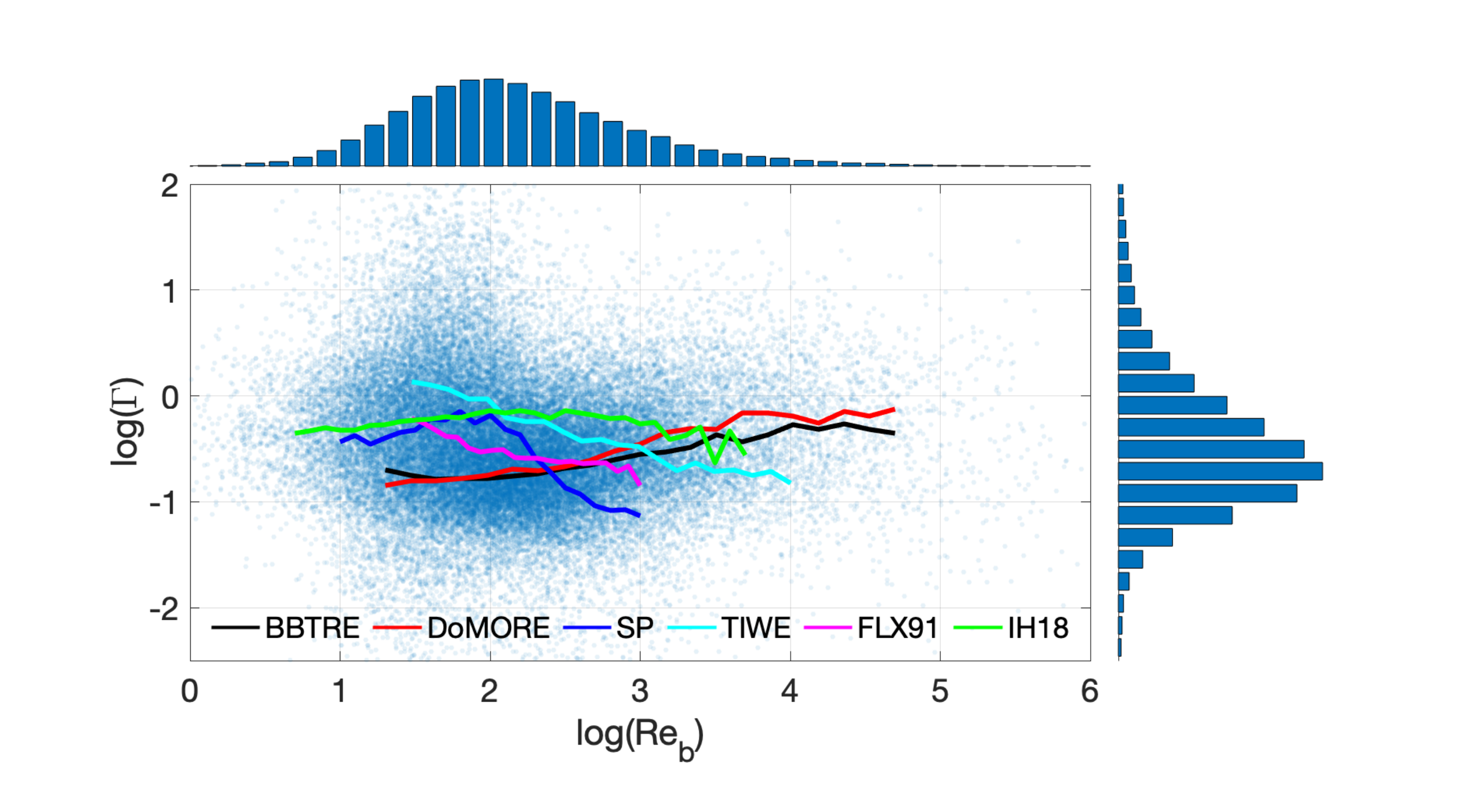}
\caption{Scatter plot of data for the six oceanic datasets in $Re_b-\Gamma$ space, overlaid by bin-averaged curves for each experiment.}
\label{Fig7}
\end{center}
\end{figure}

\color{black}
Heretofore, we have presented arguments for parameterizing the turbulent flux coefficient in terms of a ratio of length scales. On the other hand, the literature proposing the use of appropriate definitions of a buoyancy Reynolds number $Re_b$ and/or a Richardson number $Ri$ to quantify mixing efficiency is relatively well established \citep{PC03,ivey,Gregg2018MixingOcean,Caulfield2021LayeringFlows}. However, efforts that compared various datasets found them not to overlap when mapped onto these parameters~\citep{bouffard2013diapycnal,mashayek2017efficiency,monismith2018mixing}. To highlight this for the datasets employed in this article, in figure \ref{Fig7} we plot the totality of the 6 datasets considered herein in the $\Gamma-Re_b$ space and also include bin-averaged means for each dataset. Put simply, the data are all over the place. 

There is substantial empirical evidence  that in the energetically mixing regime, $\Gamma  \propto Re_b^{-1/2}$ \citep{ivey,bouffard2013diapycnal,mashayek2017efficiency,monismith2018mixing}. There is also some evidence from experimental and numerical data (see review in \cite{bouffard2013diapycnal}; also see \cite{mashayek2017efficiency}) that for smaller $Re_b$, $\Gamma \propto Re_b^{1/2}$. Finally, there is also substantial evidence for $\Gamma \propto Ri\sim Fr^{-2}$ in sufficiently weakly stratified flows when $Pr_T \sim O(1) $  \citep{Shih2005,Wells_2010,lozovatsky2013mixing,Deusebio2015,SP15,Maffioli2016MixingTurbulence,Zhou2017,Garanaik2019OnFlows}. One could argue that each line in figure \ref{Fig7} consists of left and right flanks that represent these limits. 

We \cpc{have} tried, unsuccessfully, to interpret these limits as young and decaying turbulence phases and parameterize $\Gamma$ in a fashion similar to \citet{bouffard2013diapycnal,mashayek2017efficiency} with the addition of $Ri$ to the parameterization. For that to work, most of the data would be expected to be at the transition between the left and right scaling (i.e. the peak of the lines in figure \ref{Fig7}) but that did not hold for any of the various data we considered here. From a physical perspective, the most fundamental message of this paper is the \mha{need} to account for both energy injecting and dissipation scales to quantify mixing. $Re_b$ only \cpc{conveys} information about the latter, and thus is insufficient by design. Our efforts to resort to $Ri$ as a means for including a scale for background shear and reconcile the various data was not fruitful even for the two datasets that possess $Ri$ measurements. Considering $Ri$ is not often measured, and the success of equation \eqref{G_ROT}, we did not pursue $\Gamma=f(Re_b,Ri)$ further \cpcf{here}. We also like to point out that while $Re_b$ spans five orders of magnitude in figure \ref{Fig7}, the corresponding $R_{OT}$ captures most of the data within 2 orders of magnitude. So even if a measurable energy injection scale could somehow be combined with $Re_b$,  \eqref{G_ROT} would still be much preferable, as attested to by the nice collapse of data in figure \ref{Fig3}g.

\color{black}

\section{Conclusions}\label{sec:conc}
In this paper, we have shown that a  parameterization of an appropriate definition of the turbulent flux coefficient $\Gamma$ based on the ratio $R_{OT}$ of Ozmidov and Thorpe scales may be derived using classical ideas of  energy transfer within the inertial range, mixing length theory, and criticality conditions for parallel free shear layers. The novelty of this  parameterization is that it spans the entire turbulent life cycle of a mixing event, and yet does not involve any empirical tuning coefficients. We have shown  that the parameterization agrees  well with a host of oceanic observations of turbulent overturns in different geographical, depth, and turbulence regimes. Most energetic turbulent patches appear to correspond to an intermediate phase of turbulence where $R_{OT} \sim 1$, implying an efficient transfer of energy between the background flow and the hierarchy of eddies that exist in the inertial range. We refer to this efficient and in some sense optimal phase as the `Goldilocks mixing' phase, and have shown that $\Gamma \simeq 1/3$ in this phase. This value agrees not only with observations but also directly follows from assuming a critical Richardson number $Ri_{cr} \sim 1/4$, suggesting a connection with 
such flows being in a critical `kind of equilibrium' state, or close to `marginal' stability in some sense.

\color{black}Our work further emphasizes the \cpc{essential significance} of \cpc{identification and quantification of} both the energy injecting and dissipation scales  for accurate \cpc{parameterization}. Of course, this is well known both from observations and simulations. However, it is common to infer turbulent diffusivity from microstructure measurements alone or from Thorpe scale estimates with the assumption of a fixed $R_{OT}$. We simply further emphasize that both such approaches are fundamentally limited. 
When $\epsilon$, $L_T$, and $\chi$ are all available, \eqref{eq:gammachi}, \eqref{G_ROT}, may be used in conjunction to constrain $\Gamma$ quite tightly. In the absence of $\chi$, \eqref{G_ROT} can \cpc{still} be used \cpc{successfully} to estimate $\Gamma$. 
\color{black}

We have also quantified the extent to which variations in $\Gamma$ matter for the total buoyancy flux based on the sum of contributions of individual patches for the various observational datasets, both individually and in combination,  and have shown that these variations can lead to inaccuracies  as high as a factor of 7. The key point appears to be that  the statistics of overturn characteristics have a surprisingly long and fat tail in the high $\epsilon$ limit. Since  $\epsilon$ is observed to vary over orders of magnitude, estimating $\Gamma$ correctly for the tail is crucially important, and a constant value  based on  the mean of many overturns will inevitably be inaccurate. Our parameterization describes $\Gamma$ relatively well for such tails and also demonstrates that $1/3$ may be more appropriate than $0.2$ as a rule of thumb for a   constant $\Gamma$. 

Finally, we wish to emphasize that everything that we have discussed in this paper is in relation to spatially and temporally isolated turbulent patches. Such patches are identified from profiler data that include quiescent regions. \color{black} We showed that young turbulent patches with high $\Gamma$ can bias to high values the bulk value of $\Gamma$ associated with a coarse resolution climate model grid cell. Thus, not only the high tail of $\epsilon$ matters, but so does the portion of the low tail that correspond to young patches (with the rest of the low tail corresponding to decaying turbulence)\color{black}. Outstanding  open questions, pertinent to the quantification of the role of mixing on regional and global scale in the world's oceans, are i) what is the appropriate underlying distribution of patches in time and space, and ii) what is the relative importance of the driving mechanisms of ocean turbulence. While the properties of the distribution or  `census' of the turbulent patch population and generating mechanisms are key to  any attempt to connect the physics of density-stratified turbulence to most oceanographic applications, both issues remain extremely poorly understood \citep{CM21,mackinnon2017climate}.

\subsection*{Acknowledgments}
The collection of these data took place after years of instrument development and hundreds of at-sea days, and would not have been possible without the hard work and skill of the Captain and crew of each research vessel. We would like to thank T. Ijichi and T. Hibiya for sharing the IH18 data, J. Moum for sharing the TIWE and FLX91 data, and G. Carter for sharing the SP data. DoMore and BBTRE data are publicly available through \citet{Ijichi2020HowAbyss}. 
We thank the NSF  for the  support  of the Summer Study Program in GFD at Woods Hole Oceanographic Institution, in  particular,
of  the 2019 program on Stratified Turbulence and Ocean Mixing Processes, where some of this work was carried out. We thank Lois Baker for making figure 5b based on her simulation, which were a higher resolution version of \citet{mashayek2017topographic} the simulation. We also thank B.B. Cael for making of figure 6a. Ali Mashayek acknowledges support from National Environmental Research Council (NE/P018319/1).
Declaration of Interests. The authors report no conflict of interest.
\bibliography{references}
\bibliographystyle{jfm}

\end{document}